\documentclass[lettersize, journal]{IEEEtran}
\IEEEoverridecommandlockouts
\usepackage{cite}
\usepackage{amsmath,amsthm}
\usepackage{amsfonts}
\usepackage{array}
\usepackage{amssymb}
\usepackage{caption}
\usepackage[font=small]{caption}
\usepackage{graphicx}
\usepackage{stfloats}
\usepackage[labelformat=simple]{subcaption}
\usepackage{hyperref}
\hypersetup{hidelinks}
\usepackage{textcomp}
\usepackage{booktabs}
\usepackage[ruled,vlined]{algorithm2e}
\usepackage{algorithmic}
\def\BibTeX{{\rm B\kern-.05em{\sc i\kern-.025em b}\kern-.08em
    T\kern-.1667em\lower.7ex\hbox{E}\kern-.125emX}}

\ifCLASSINFOpdf

\usepackage{xcolor}
\usepackage{tcolorbox}

\usepackage{acro}

\newtheorem{assump}{Assumption}
\usepackage{tabu,longtable}

\begin{document}
\begin{sloppypar}

\title{Scalable Near-Field Localization Based on Partitioned Large-Scale Antenna Array
}
\author{Xiaojun Yuan,~\IEEEmembership{Senior~Member,~IEEE,} 
Yuqing Zheng,~\IEEEmembership{Graduate~Student~Member,~IEEE,}
Mingchen Zhang,~\IEEEmembership{Graduate~Student~Member,~IEEE,}
Boyu Teng,~\IEEEmembership{Graduate~Student~Member,~IEEE,}
and Wenjun Jiang,~\IEEEmembership{Graduate~Student~Member,~IEEE}
\thanks{X. Yuan, Y. Zheng, M. Zhang, B. Teng, and W. Jiang are with the National Key Laboratory of Wireless Communications, University of Electronic Science and Technology of China, Chengdu 611731, China (e-mail: xjyuan@uestc.edu.cn; \{yqzheng, zhangmingchen, byteng, wjjiang\}@std.uestc.edu.cn). This paper was presented in part at the 2024 IEEE International Conference on Communications, Denver, Colorado, Jun. 2024 \cite{zheng_2024_ICC}.}
}
\maketitle
\begin{abstract}
This paper studies a passive localization system, where an extremely large-scale antenna array (ELAA) is deployed at the base station (BS) to locate a user equipment (UE) residing in its near-field (Fresnel) region. We propose a novel algorithm, named array partitioning-based location estimation (APLE), for scalable near-field localization. The APLE algorithm is developed based on the basic assumption that, by partitioning the ELAA into multiple subarrays, the UE can be approximated as in the far-field region of each subarray. We establish a Bayeian inference framework based on the geometric constraints between the UE location and the angles of arrivals (AoAs) at different subarrays. Then, the  APLE algorithm is designed based on the message-passing principle for the localization of the UE. APLE exhibits linear computational complexity with the number of BS antennas, leading to a significant reduction in complexity compared to existing methods. We further propose an enhanced APLE (E-APLE) algorithm that refines the location estimate obtained from APLE by following the maximum likelihood principle. The E-APLE algorithm achieves superior localization accuracy compared to APLE while maintaining a linear complexity with the number of BS antennas. Numerical results demonstrate that the proposed APLE and E-APLE algorithms outperform the existing baselines in terms of localization accuracy. 

\end{abstract}
\begin{IEEEkeywords}
    Near-field localization, extremely large-scale antenna array, array partitioning, Bayesian inference
\end{IEEEkeywords}
\section{Introduction}\label{Intro}
\par Integrated sensing and communication (ISAC) has emerged as a highly promising technology for sixth-generation (6G) mobile communications \cite{liuIntegratedSensingCommunications2022,liuSurveyFundamentalLimits2022}. This is driven by the growing demand for high-quality communication and sensing services in various emerging 6G application scenarios, such as augmented reality (AR), internet of vehicles (IoV), and unmanned aerial vehicle (UAV) communications. In these applications, efficient and precise acquisition of user equipment's (UE's) location information is of critical importance, with centimeter-level localization accuracy expected for ensuring required service quality\cite{6G9040431}. Much research effort has been devoted to leveraging other emerging technologies, such as Terahertz communication, intelligent reflecting surface (IRS), and extremely large multi-input-multi-output (XL-MIMO) to provide enhanced localization services in 6G\cite{Teng9772371, MCRB}.
\par Among these new technologies, XL-MIMO is envisioned to have the potential to greatly improve the localization capabilities of wireless networks\cite{XLMIMOLoc}. To meet the demands for higher spectral efficiency and system capacity in 6G networks, the utilization of larger-scale antenna arrays has become a realistic trend\cite{ELAAtrendarXiv230806455Q}.  
Compared with fifth-generation (5G) massive MIMO which involves up to hundreds of antennas deployed at a base station (BS), 6G XL-MIMO is expected to employ extremely large-scale antenna arrays (ELAA) comprising thousands or even tens of thousands of antennas\cite{ATutorial2023arXiv230707340W}. The deployment of ELAA enables anchor points (typically, BSs) to collect enough measurements of the targets even in one snapshot, thereby achieving high accuracy and robustness of wireless localization\cite{BJORNSON20193}.
\par There are, however, many challenging issues to be addressed before wireless localization can reap the full benefits of XL-MIMO. First of all, traditional localization problems based on the far-field assumption typically only involve estimating the target's directions, referred to as AoA estimation. However, in XL-MIMO systems, due to the expanded aperture of the BS ELAA and the employment of a higher frequency band, targets are more likely to be located in the near-field region (also known as the Fresnel region) of the BS ELAA. Consequently, accurate target localization now requires the simultaneous estimation of both a target's direction and its distance from the BS. Near-field localization has been studied in the past few years. Approaches to near-field localization, such as those based on  time-of-arrival (TOA)\cite{ToA9625826}, time difference of arrival (TDoA)\cite{TDOA8666171}, and fingerprinting\cite{finger9773170}, have been studied. 
\par 

The existing near-field localization algorithms, however, generally suffer from scalability issues when applied to XL-MIMO scenarios. Take the well-known multiple signal classification (MUSIC) based algorithm \cite{MUSICFast10022701} as an example. The complexity of the MUSIC-based algorithm primarily arises from the eigen decomposition of the received signal correlation matrix, where the complexity increases cubically with the number of antennas. In the case of ELAA, where the number of antennas can reach thousands or even tens of thousands, this cubic complexity leads to a prohibitively high computational burden for the receiver, thereby being incapable of providing real-time localization services. Other high-precision near-field localization algorithms, such as those based on ESPRIT and compressed sensing\cite{ESPRIT229094735, CSNFC9709801}, also exhibit cubic or even higher complexity with the number of BS antennas. 

In addition to the scalability issue, the localization accuracy of the existing methods is also unsatisfactory \cite{CSNFC9709801, Fresref1, Fresref2, Fresref3}. These methods mostly rely on the Fresnel approximation \cite{Fresappro} to simplify the channel model. The symmetry of the BS array is utilized to construct a special correlation matrix of the received signals, which decouples the direction and distance information of a target. The location of the target is then recovered by separately estimating the direction and distance of the target relative to the BS. Clearly, the introduction of the Fresnel approximation may compromise the localization accuracy of these methods. Moreover, these approaches often impose a limitation on the antenna spacing at the BS array, typically requiring the spacing to be less than a quarter of the wavelength. When applied to commonly used half-wavelength spaced arrays, these methods suffer from severe performance degradation due to phase ambiguity. As such, it is of crucial importance to develop accurate yet scalable near-field localization algorithms for XL-MIMO systems. 

\par To tackle the above issues, we propose a novel algorithm for scalable near-field localization, named \textit{array partitioning-based location estimation (APLE) algorithm}. Specifically, we consider a passive localization system, where an ELAA is deployed at the BS to locate a single UE in its near-field region using the received signals. 
We assume that by partitioning the ELAA into multiple subarrays, the UE can be approximated as in the far-field region of each subarray. Owing to a distinct relative location between each subarray and the UE, the signal transmitted by the UE exhibits varying AoA as they arrive at different subarrays, known as \textit{AoA drifting}. The APLE algorithm leverages the AoA drifting effect for UE localization. A posterior probability model of the UE location is established based on the received signal model and the geometric constraints between the UE location and the observed AoAs. Message passing is performed based on a factor graph representation of this posterior probability model. For messages difficult to compute, we introduce approximate calculation methods. The proposed APLE algorithm exhibits a \textit{linear} complexity with the number of BS antennas, which is a significant reduction compared to the common cubic complexity of the existing near-field localization algorithms. Numerical results demonstrate that the estimation-error performance achieved by APLE greatly outperforms that of other baseline methods, and can approach a misspecified Cramér-Rao bound (MCRB) for the considered near-field localization problem.

To improve the localization accuracy, we further propose an enhanced version of APLE, namely the enhanced APLE (E-APLE) algorithm. E-APLE refines the location estimate of APLE by following the maximum likelihood (ML) principle. Specifically, under the near-field signal model of the entire BS array, we formulate the ML problem for the UE location. We show that the log-likelihood function of the UE location generally appears in a ridge shape in the distance-angle polar domain. Inspired by this ridge-shaped feature, we employ a block coordinate ascent (BCA) method to solve the ML problem by iteratively updating the distance and angle parameters. 
Furthermore, we show that the log-likelihood function is highly multi-modal. To prevent the BCA algorithm from becoming trapped in poor local maxima, we initialize the BCA algorithm by the promising estimate of the UE location obtained from APLE. Numerical results demonstrate that E-APLE achieves superior localization accuracy compared to APLE, and can closely approach the Cramér-Rao bound (CRB) for the considered near-field localization problem. 

The contributions of this paper are summarized as follows:
\begin{itemize}
\item[1)] We introduce the notion of array partitioning for near-field localization. Specifically, we establish a subarray far-field signal model based on a basic assumption, that is, with an appropriate array partitioning strategy, a UE in the near-field region of the entire BS array can be in the far-field region of each subarray. By exploiting the geometric relationships between the UE location and the AoAs at different subarrays, we formulate a probabilistic near-field localization problem under the Bayesian inference framework.
\item[2)] We propose the APLE algorithm for solving this probabilistic near-field localization problem. APLE operates by performing message passing on a factor graph representation of the subarray far-field signal model. To handle computationally challenging messages, we propose approximate calculation methods. The complexity of APLE scales linear with the number of BS antennas.
\item[3)] We further propose the E-APLE algorithm. The idea of E-APLE is to refine the location estimate obtained from APLE by following the ML principle. E-APLE offers superior localization accuracy compared to APLE while maintaining a linear complexity with the number of BS antennas.
\item[4)] We derive the CRB and MCRB for the considered near-field localization problem. 
\item[5)] We show by numerical results that the proposed APLE and E-APLE algorithms significantly outperform the baseline methods and meanwhile exhibit much lower runtimes. Besides, at high signal-to-noise ratio (SNR), the estimation-error performance of the E-APLE algorithm can closely approach the CRB.
\end{itemize}

\par
The remainder of this paper is organized as follows. Section \ref{S2} introduces the near-field ELAA localization system. The array partitioning and probabilistic problem formulation are discussed in Section \ref{S3}. In Section \ref{S4}, we introduce the message-passing-based APLE algorithm. In Section \ref{S5}, we develop the E-APLE algorithm. CRB and MCRB of the considered localization problem are derived in \ref{S6}. Numerical results are presented in Section \ref{S7}, and the paper is concluded in Section \ref{S8}.
\par
\textit{Notations:} We use lower-case and upper-case bold letters to denote vectors and matrices, respectively. We use $(\cdot)^{\mathrm{T}}$ and $(\cdot)^{\mathrm{H}}$ to denote the operations of transpose and conjugate transpose, respectively. We use $\text{Re}[\mathbf{X}]$ to denote the real part of $\mathbf{X}$, $\left[\cdot \right]_{i,j}$ to denote the element in the $i$-th row, $j$-th column of a matrix. We use $\mathcal{N}\left(\boldsymbol{x};\boldsymbol{\mu},\boldsymbol{\Sigma}\right)$ and $\mathcal{CN}\left(\boldsymbol{x};\boldsymbol{\mu},\boldsymbol{\Sigma}\right)$ to denote the Gaussian distribution and the circularly-symmetric Gaussian distribution with mean vector $\boldsymbol{\mu}$ and covariance matrix $\boldsymbol{\Sigma}$, $\mathcal{M}({x};{\mu},{\kappa})$ to denote the Von Mises (VM) distribution with mean direction ${\mu}$ and concentration parameters ${\kappa}$. We use $\mathbb{E}[\cdot]$ to denote the expectation operator, $\times$ to denote the cross product, $\odot$ to denote the Hadamard product, $\otimes$ to denote the Kronecker product, $\partial$ to denote the partial derivative operator, $\delta(\cdot)$ to denote the Dirac delta function, $\|\cdot\|$ to denote the $\ell_{2}$ norm, and $\jmath$ to denote the imaginary unit. $\mathbf{I}_N$ denotes the $N \times N$ identity matrix and $\mathcal{I}_N$ represents the index set $\{1,\ldots,N\}$, where $N$ is a positive integer.
\begin{figure}[t]
    \centering
    \includegraphics[width=.8\linewidth]{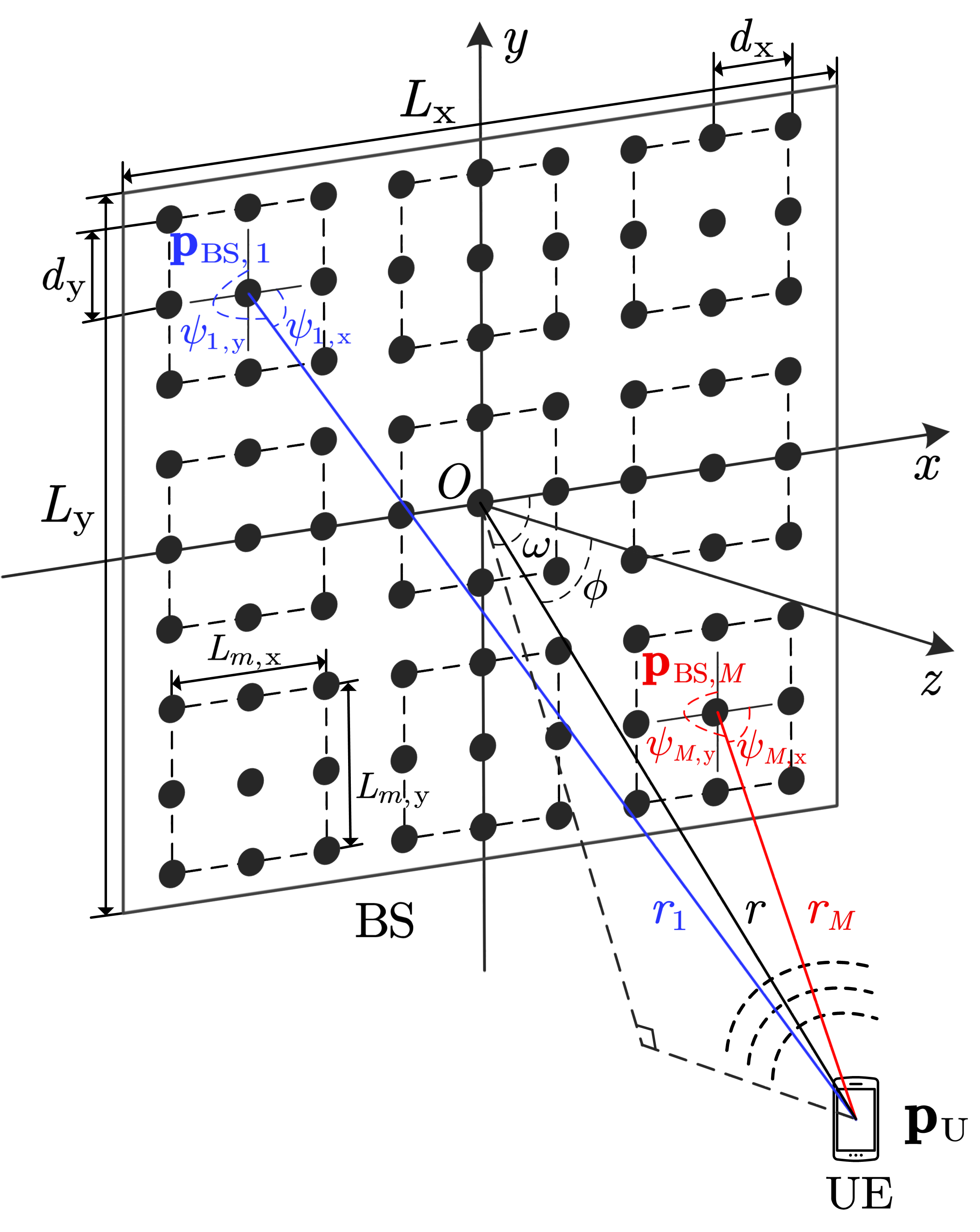}
    \caption{The uplink communication system with the UE located in the near-field region of the BS array.}
    \label{system}
\end{figure}

\section{System Model}\label{S2}
As illustrated in Fig. \ref{system}, we consider an uplink communication system consisting of one BS and a single UE. The BS is equipped with an $N_{\mathrm{B}}$-antenna ELAA arranged as a uniform planar array (UPA), and the UE is equipped with a single antenna. A 3D Cartesian coordinate system is established with the center of the BS array located at the origin $O$. The $x$-axis and the $y$-axis are parallel to the two sides of the BS array, and the $z$-axis is perpendicular to the BS array. We assume  $N_{\mathrm{B}}=N_{\mathrm{x}} \times N_{\mathrm{y}} $, where $N_{\mathrm{x}}$ and $N_{\mathrm{y}}$ are the number of antennas along the $x$-axis and $y$-axis, respectively. The uniform antenna spacings of the BS array along the $x$-axis and $y$-axis are denoted by $d_{\mathrm{x}}$ and $d_{\mathrm{y}}$, respectively. The size of the BS array is given by $L_{\mathrm{x}} \times L_{\mathrm{y}}$, where $L_{\mathrm{x}} = N_{\mathrm{x}} d_{\mathrm{x}}$ and $L_{\mathrm{y}} = N_{\mathrm{y}} d_{\mathrm{y}}$. 
Denote by $\mathbf{p}_{\mathrm{BS},(i,j)}=\left[\left(i-\frac{N_{\mathrm{x}}+1}{2} \right)d_{\mathrm{x}}, \left(j-\frac{N_{\mathrm{y}}+1}{2} \right)d_{\mathrm{y}}, 0 \right]^\mathrm{T}$  the location of the $(i,j)$-th antenna at the BS array, $i\in \mathcal{I}_{N_{\mathrm{x}}}$, $j\in \mathcal{I}_{N_{\mathrm{y}}}$. Deonte by $\mathbf{p}_{\mathrm{U}}=[x,y,z]^\mathrm{T}$ the location of the UE, and by $r=\left\|\mathbf{p}_{\mathrm{U}} \right\|$ the link distance between the UE and the center of the BS array. 

\par 
ELAAs are typically deployed in millimeter-wave
(mmWave) and terahertz (THz) frequency bands, where the channel strength disparity between line-of-sight (LoS) and non-line-of-sight (NLoS) links can easily exceed 20 dB\cite{NLoS}. Hence, we only consider modeling the LoS channel between the BS array and the UE. Denote by $D\triangleq \left(L_{\mathrm{x}}^2+L_{\mathrm{y}}^2\right)^{\frac{1}{2}}$ the largest dimension of the BS array. We assume that the UE is located outside the reactive near-field region of the BS array, i.e., $R_{\mathrm{FS}}<r$, where $R_{\mathrm{FS}} \triangleq \sqrt[3]{\frac{D^4}{8\lambda}}$ represents the Fresnel distance of the BS array \cite{Fraunhofer_Selvan_7942128}. Besides, we neglect the variations in the channel path loss between the UE and different BS antennas. The distance between the UE and the $(i,j)$-th antenna of the BS array is given by $ r_{(i,j)}=\left\|\mathbf{p}_{\mathrm{U}}-\mathbf{p}_{\mathrm{BS},(i,j)}\right\|$, $i \in \mathcal{I}_{N_{\mathrm{x}}}, j \in \mathcal{I}_{N_{\mathrm{y}}}$. Denote by $\lambda$ the carrier wavelength, and by $\beta$ the common path loss coefficient. The channel coefficient between the UE and the $(i,j)$-th antenna of the BS array is expressed as
\begin{equation}\label{h}
 	h_{(i,j)}=\beta e^{-\jmath\frac{ 2\pi}{\lambda} r_{(i,j)}},~ i \in \mathcal{I}_{N_{\mathrm{x}}},~ j \in \mathcal{I}_{N_{\mathrm{y}}}.
\end{equation}
Denote by $s$ the signal transmitted from the UE. The received one-snapshot signal at the BS array can be expressed as 
\begin{equation}\label{eq1}
\boldsymbol{y}=\boldsymbol{h} s+\boldsymbol{n},
\end{equation}
where $\boldsymbol{h}\in\mathbb{C}^{N_{\mathrm{B}}}$ represents the channel vector between the UE and the BS array, with the $\left(\left(i-1\right) N_{\mathrm{y}}+j\right)$-th element of $\boldsymbol{h}$ given by \eqref{h}, $i\in \mathcal{I}_{N_{\mathrm{x}}}, j \in \mathcal{I}_{N_{\mathrm{y}}}$; $\boldsymbol{n}\in\mathbb{C}^{N_{\mathrm{B}} }$ denotes the circularly symmetric complex Gaussian noise that follows $ \mathcal{C N}\left( \mathbf{0},\sigma^{2} \mathbf{I}_{N_{\mathrm{B}}}\right)$, where $\sigma^{2}$ is the noise variance. With the equivalent complex channel gain defined by $\alpha \triangleq \beta s$, the received signal can be rewritten as 
\begin{equation}\label{ENFM}
  \boldsymbol{y}=\alpha \mathbf{a}\left(\mathbf{p}_{\mathrm{U}}\right)+\boldsymbol{n},  
\end{equation}
where $\mathbf{a}\left(\mathbf{p}_{\mathrm{U}}\right)\in\mathbb{C}^{N_{\mathrm{B}}}$ represents the near-field steering vector of the BS array, with the $\left(\left(i-1\right) N_{\mathrm{y}}+j\right)$-th element of $\mathbf{a}(\mathbf{p}_{\mathrm{U}})$ being $e^{-\jmath\frac{ 2\pi}{\lambda} r_{(i,j)}}$, $i\in \mathcal{I}_{N_{\mathrm{x}}}, j \in \mathcal{I}_{N_{\mathrm{y}}}$. The SNR is defined as ${\left| \alpha \right|^2}/{\sigma^2}$.

\par We assume that the BS location $\{\mathbf{p}_{\mathrm{BS},(i,j)}|i \in \mathcal{I}_{N_{\mathrm{x}}}, j \in \mathcal{I}_{N_{\mathrm{y}}} \}$ is known. Our goal is to estimate the UE location $\mathbf{p}_{\mathrm{U}}$
based on the received signal $\boldsymbol{y}$. Based on the received signal model in \eqref{ENFM}, the probability density function (pdf) of $\boldsymbol{y}$ conditioned on $ \mathbf{p}_{\mathrm{U}}$ and $\alpha $ is given by
\begin{equation}\label{nearMLE}
p(\boldsymbol{y}| \mathbf{p}_{\mathrm{U}}, \alpha  )= \mathcal{CN}\left(\boldsymbol{y}; \alpha \mathbf{a}(\mathbf{p}_{\mathrm{U}}),\sigma^2\mathbf{I}_{N_{\mathrm{B}}}\right),
\end{equation}
with the log-likelihood function given by 
\begin{align}\label{llf}
 \ln p(\boldsymbol{y};\mathbf{p}_{\mathrm{U}},\alpha)\propto -\frac{1}{\sigma^2} \vert|\boldsymbol{y} - \alpha \mathbf{a}(\mathbf{p}_{\mathrm{U}}) \vert|^2.
\end{align}

The log-likelihood function $\ln p(\boldsymbol{y};\mathbf{p}_{\mathrm{U}},\alpha)$ is highly multi-modal, especially when the UE is located close to the BS. In this case, the straightforward gradient ascent (GA) methods may struggle to find the global optima of the log-likelihood function. Performing a 3D exhaustive search over all possible values of $\mathbf{p}_{\mathrm{U}}$ can be used to solve this ML problem, but this approach is notoriously time-consuming.

\begin{figure}[t]
	\centering
	\begin{subfigure}{0.8\linewidth}
		\centering
		\includegraphics[width=1\linewidth]{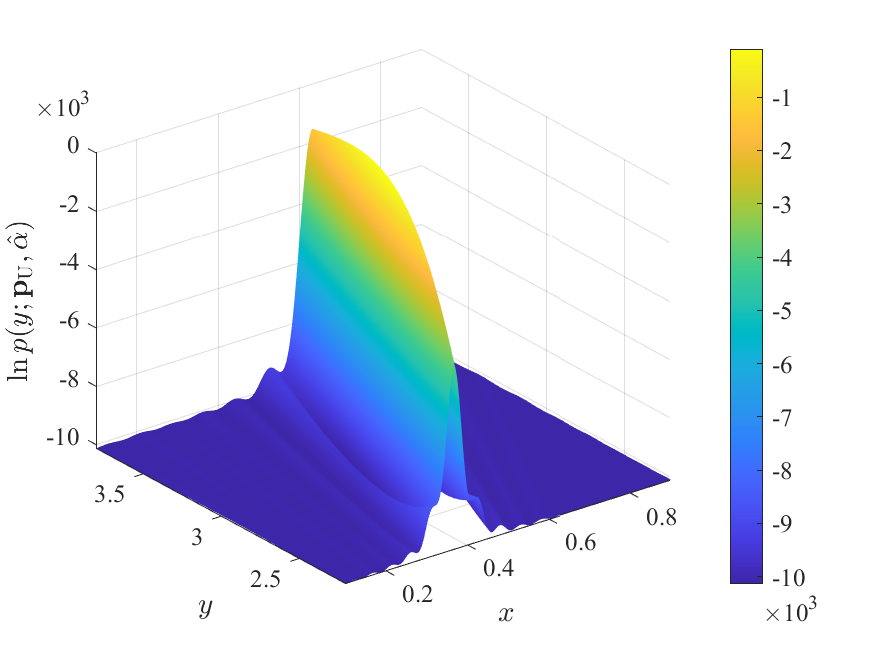}
		\caption{$N_{\mathrm{B}}=100$}
		\label{MLE100}
	\end{subfigure}
	\centering
	\begin{subfigure}{0.8\linewidth}
		\centering
		\includegraphics[width=1\linewidth]{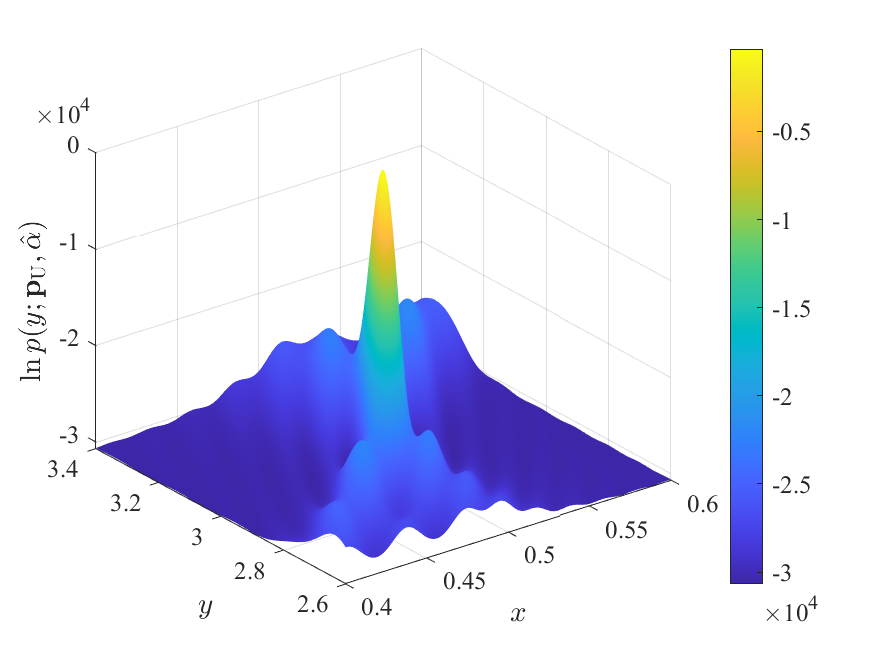}
		\caption{$N_{\mathrm{B}}=300$}
		\label{MLE300}
	\end{subfigure}
	\centering
	\begin{subfigure}{0.8\linewidth}
		\centering
		\includegraphics[width=1\linewidth]{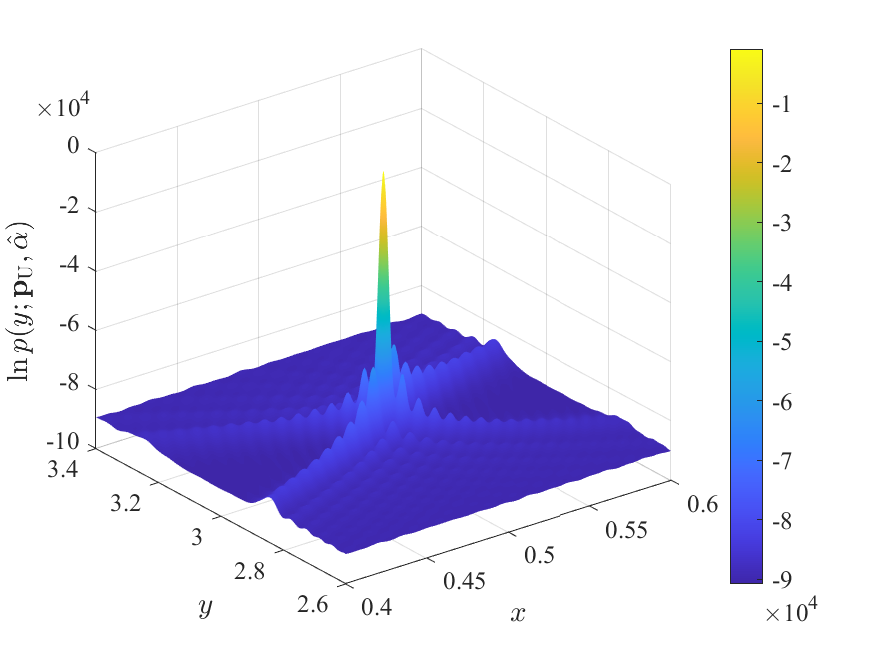}
		\caption{$N_{\mathrm{B}}=900$}
		\label{MLE900}
	\end{subfigure}
	\caption{  The log-likelihood function $\ln p(\boldsymbol{y};\mathbf{p}_{\mathrm{U}},\hat{\alpha})$ with different array sizes. $N_{\mathrm{y}}=1$, $N_{\mathrm{x}}=N_{\mathrm{B}}$, SNR $=20$ dB, $\lambda=0.01$ m, $d_{\mathrm{x}} = \frac{\lambda}{2}$, ${\mathbf{p}}_{\mathrm{U}}=\left[ x,y\right]^{\mathrm{T}}=[0.5,3]^{\mathrm{T}}$.} 
	\label{FigMLE}
\end{figure}

\par For illustration, we consider a reduced two-dimensional case, where a UE is located in the near-field region of a uniform linear array (ULA), i.e., $N_{\mathrm{y}}=1$. The UE location ${\mathbf{p}}_{\mathrm{U}}$ for this toy two-dimensional case is expressed as ${\mathbf{p}}_{\mathrm{U}}=\left[ x,y\right]^{\mathrm{T}}$. 
For the log-likelihood function $\ln p(\boldsymbol{y};\mathbf{p}_{\mathrm{U}},{\alpha})$, we replace $\alpha$ by the least-square estimate $\hat{\alpha} = \frac{\mathbf{a}^{\mathrm{H}}({\mathbf{p}}_{\mathrm{U}})\boldsymbol{y}}{\vert| \mathbf{a}({\mathbf{p}}_{\mathrm{U}})\vert|^2}$; see Section \ref{S5} for details.
The log-likelihood function $\ln p(\boldsymbol{y};\mathbf{p}_{\mathrm{U}},\hat{\alpha})$ with different array sizes is shown in Fig. \ref{FigMLE}. It can be observed that as the array size increases, the peak of the log-likelihood function becomes sharper. This means that a larger array size can significantly enhance the estimation accuracy of the localization problem. However, we also observe that the log-likelihood function is highly multi-modal, and the number of local maximum points increases with the enlargement of the array size. For a large $N_{\mathrm{B}}$ (e.g., $N_{\mathrm{B}}=900$), conventional GA-based methods for finding $\mathbf{p}_{\mathrm{U}}$ can be easily trapped in local maxima, and an exhaustive search over $x$ and $y$ is very costly. Even a slight deviation in the estimate of $x$ can lead to the failure of searching $y$. This shows that finding the global peak of the likelihood function \eqref{llf} is very challenging in an ELAA scenario.
\par To address the above issue, we next propose an efficient near-field localization method based on array partitioning. In this method, the BS array is partitioned into multiple subarrays to ensure that the user is located in the far-field region of each subarray. By adopting a subarray far-field signal model, we formulate the near-field localization problem under the Bayesian inference framework. The key challenge lies in estimating the UE location by appropriately combining the AoA estimates from various subarrays with subtle differences. The proposed method circumvents the challenges associated with the direct maximum likelihood methods. Details of the proposed array partitioning approach are provided in subsequent sections.

\section{Array Partitioning Based Problem Reformulation}\label{S3}

\subsection{Array Partitioning and Subarray Far-Field Assumption}
\par We partition the BS array into $M$ non-overlapping subarrays as shown in Fig. \ref{system}. The $m$-th subarray consists of $N_{m}=N_{m,\mathrm{x}} \times N_{m,\mathrm{y}}$ antennas, where $N_{m,\mathrm{x}}$ and $N_{m,\mathrm{y}}$ are the number of antennas in the $m$-th subarray along the $x$-axis and $y$-axis, respectively, $m\in \mathcal{I}_{M}$. The size of each subarray is denoted by $L_{m,\mathrm{x}} \times L_{m,\mathrm{y}}$ with $L_{m,\mathrm{x}} = N_{m,\mathrm{x}} d_{\mathrm{x}}$ and $L_{m,\mathrm{y}} = N_{m,\mathrm{y}} d_{\mathrm{y}}$. Thus the largest dimension of the $m$-th subarray is denoted by $D_{m} = \left(L_{m,\mathrm{x}}^2+L_{m,\mathrm{y}}^2\right)^{\frac{1}{2}}$. The Fraunhofer distance of the $m$-th subarray is given by $R_{m, \mathrm{FH}}= \frac{2D_{m}^{2}}{\lambda}$. 
The location of the center  of the $m$-th subarray is denoted by $\mathbf{p}_{\mathrm{BS},m}\in\mathbb{R}^{3}$. The link distance between the UE and the center of the $m$-th subarray is denoted by ${r}_{m} =\left\|\mathbf{p}_{\mathrm{U}}- \mathbf{p}_{\mathrm{BS},m}\right\|$.
We make the following \textit{subarray far-field assumption (SFA)}.
\begin{assump}\label{SubArrAssump} (\textbf{subarray far-field assumption}):
    With a sufficiently large $M$ and an appropriate partition strategy, the UE is located in the far-field region of each BS subarrays, i.e., $R_{m,\mathrm{F}}<r_m$, $\forall m \in \mathcal{I}_M$.
\end{assump}
Assumption \ref{SubArrAssump} means that the UE can be located in the near-field region of the entire BS array meanwhile in the far-field region of each subarray. 
For a simple justification, consider a BS array with dimensions $L_{\mathrm{x}}=L_{\mathrm{y}}=0.5$ m and a wavelength of $\lambda = 0.01$ m. The BS array is partitioned into $M=25$ subarrays with $L_{m,\mathrm{x}}=L_{m,\mathrm{y}}=0.1$ m. In this case, the Fresnel distance and the Fraunhofer distance of the BS array are $R_{\mathrm{FS}}=0.17$ m and $R_{\mathrm{FH}}=100$ m, respectively. The Fraunhofer distance of each subarray is $R_{m, \mathrm{FH}}=4$ m, $m\in \mathcal{I}_{M}$.
There is a large overlap between the near-field region of the entire BS array (i.e., $R_{\mathrm{FS}}<r_m<R_{\mathrm{FH}}$) and the far-field region of each subarray (i.e., $R_{m,\mathrm{FH}}<r_m$). Next, we introduce the simplified received signal model based on the SFA.

\subsection{Subarray Far-Field Model}
\par Take the center of the $m$-th subarray as the reference point. Let $\Tilde{k} \triangleq \left( k-\frac{ N_{m,{\mathrm{x}}} +1 }{2} \right)$ and $\Tilde{l}\triangleq \left(l-\frac{ N_{m,{\mathrm{y}}} +1 }{2} \right)$, for $k\in \mathcal{I}_{N_{m,\mathrm{x}}}, l \in \mathcal{I}_{N_{m,\mathrm{y}}}$. The location of the $(k,l)$-th antenna at the $m$-th subarray is denoted by
\begin{equation}
 \mathbf{p}_{m,(k,l)}= \mathbf{p}_{\mathrm{BS},m} + \left[ \Tilde{k}d_{\mathrm{x}}, \Tilde{l}d_{\mathrm{y}}, 0 \right]^\mathrm{T},   
\end{equation}
for $k\in \mathcal{I}_{N_{m,\mathrm{x}}}, l \in \mathcal{I}_{N_{m,\mathrm{y}}}$. As shown in Fig. \ref{system}, the angles between vector $\mathbf{p}_{\mathrm{U}}- \mathbf{p}_{\mathrm{BS},m}$ and the $x$-axis and $y$-axis are denoted by $\psi_{m,\mathrm{x}}$ and $\psi_{m,\mathrm{y}}$, respectively. Then, the cosine values $\theta_{{m,\mathrm{x}}}\triangleq\cos{\psi_{m,\mathrm{x}}}$ and $\theta_{{m,\mathrm{y}}}\triangleq\cos{\psi_{m,\mathrm{y}}}$ are respectively expressed as
\begin{align}
     \theta_{{m,\mathrm{x}}}&= \frac{(\mathbf{p}_{\mathrm{U}}-\mathbf{p}_{\mathrm{BS},m})^\mathrm{T}\mathbf{e}_\mathrm{x}}{\left\|\mathbf{p}_{\mathrm{U}}-\mathbf{p}_{\mathrm{BS},m}\right\|} \in [-1,1],  \\
     \theta_{{m,\mathrm{y}}}&= \frac{(\mathbf{p}_{\mathrm{U}}-\mathbf{p}_{\mathrm{BS},m})^\mathrm{T}\mathbf{e}_\mathrm{y}}{\left\|\mathbf{p}_{\mathrm{U}}-\mathbf{p}_{\mathrm{BS},m}\right\|} \in [-1,1].
\end{align}   
Denote by $\mathbf{u}_m = \left[\theta_{{m,\mathrm{x}}}, \theta_{{m,\mathrm{y}}}, \sqrt{1-\theta^2_{{m,\mathrm{x}}}-\theta^2_{{m,\mathrm{y}}} } \;\right]$ the unit direction vector of $\mathbf{p}_{\mathrm{U}}-\mathbf{p}_{\mathrm{BS},m}$.
The link distance between the UE and the $(k,l)$-th antenna is given by $r_{m,(k,l)} = \left \|\mathbf{p}_{m,(k,l)}-\mathbf{p}_{\mathrm{BS},m}-r_m \mathbf{u}_m\right \|$, $m\in \mathcal{I}_M$, $k\in \mathcal{I}_{N_{m,\mathrm{x}}}$, $l \in \mathcal{I}_{N_{m,\mathrm{y}}}$.
 
 \par Based on \eqref{h}, the channel coefficient between the UE and the $(k,l)$-th antenna of the $m$-th subarray is expressed as
\begin{equation}\label{hm}
         h_{m,(k,l)}=h_m e^{-\jmath\frac{ 2\pi}{\lambda} (r_{m,(k,l)} - r_m)},
\end{equation}
where $h_m =\beta e^{-\jmath\frac{ 2\pi}{\lambda}{r_m}}$ is the channel coefficient between the UE and the reference point of the $m$-th subarray, $m\in \mathcal{I}_M$, $k\in \mathcal{I}_{N_{m,\mathrm{x}}}, l \in \mathcal{I}_{N_{m,\mathrm{y}}}$.
 Under the SFA, the link distance between the UE and the $(k,l)$-th antenna in \eqref{hm} can be approximated by
\begin{subequations}
\begin{align}
 r_{m,(k,l)} &= \left \|\mathbf{p}_{m,(k,l)}-\mathbf{p}_{\mathrm{BS},m}-r_m \mathbf{u}_m\right \| \label{8a}  \\
 & = r_m - \Tilde{k} d_{\mathrm{x}}\theta_{m,\mathrm{x}}-\Tilde{l}d_{\mathrm{y}}\theta_{m,\mathrm{y}},\label{8b}
\end{align} 
\end{subequations}
where \eqref{8b} is a first-order Taylor expansion of \eqref{8a} at $\frac{\|\mathbf{p}_{m,(k,l)}-\mathbf{p}_{\mathrm{BS},m} \|}{r_m}=0$. Then, the channel coefficient $h_{m,(k,l)}$ in \eqref{hm} can be rewritten as
\begin{equation}\label{hm_2}
      \   h_{m,(k,l)} = h_m e^ {\jmath \frac{ 2\pi}{\lambda} \left( \Tilde{k}d_{\mathrm{x}}\theta_{m,\mathrm{x}}+ \Tilde{l}d_{\mathrm{y}}\theta_{m,\mathrm{y}} \right)}. 
\end{equation}
Based on \eqref{ENFM} and \eqref{hm_2}, the received signal model at the $m$-th subarray can be simplified as
\begin{align}\label{far}
    \boldsymbol{y}_{m,\mathrm{F}}&=  \alpha_{m}\mathbf{a}_{\mathrm{F}}(\theta_{m,\mathrm{x}},\theta_{m,\mathrm{y}})+\boldsymbol{n}_{m},
\end{align}
where $\alpha_{m}=h_m s$ is the equivalent channel coefficient of the $m$-th subarray. $\mathbf{a}_{\mathrm{F}}(\theta_{m,\mathrm{x}},\theta_{m,\mathrm{y}})=\mathbf{a}_\mathrm{x,F}(\theta_{m,\mathrm{x}})\otimes\mathbf{a}_\mathrm{y,F}(\theta_{m,\mathrm{y}})$ is the two-dimensional far-field steering vector of the $m$-th subarray, where $\otimes$ is the Kronecker product and $\mathbf{a}_{u,\mathrm{F}}(\theta_{m,u}) = [\mathrm{e}^{-\jmath\pi(N_{m,u}-1)d_{u}\theta_{m,u}/{\lambda }},\dots,\mathrm{e}^{{\jmath\pi(N_{m,u}-1)}d_{u}\theta_{m,u}/{\lambda }}]^\mathrm{T}$, for $u\in \{\mathrm{x},\mathrm{y}\}$. Denote by $\boldsymbol{n}_{m}\in\mathbb{C}^{N_m }$ the circularly symmetric complex Gaussian noise that follows $\mathcal{C N}\left( \mathbf{0}, \sigma^{2} \mathbf{I}_{N_{m}}\right) $. Eq. \eqref{far} is referred to as the \textit{subarray far-field model (SFM)} under the SFA.

\subsection{Probabilistic Problem Formulation}
\par We now establish the probability model of the localization problem. Based on the SFM in \eqref{far}, for $m\in \mathcal{I}_{M}$, the likelihood function of $\theta_{m,\mathrm{x}}$, $\theta_{m,\mathrm{y}}$ and $\alpha_{m}$ given $\boldsymbol{y}_{m,\mathrm{F}}$ is
\begin{multline}
p(\boldsymbol{y}_{m,\mathrm{F}}|\theta_{m,\mathrm{x}},\theta_{m,\mathrm{y}},\alpha_{m})=\\ \mathcal{CN}\left(\boldsymbol{y}_{m,\mathrm{F}}; \alpha_{m} \mathbf{a}_m(\theta_{m,\mathrm{x}},\theta_{m,\mathrm{y}}),\sigma^2\mathbf{I}_{N_{m}}\right).   
\end{multline}
Under the geometric constraints of the UE location and the $m$-th subarray, the conditional pdf $p(\theta_{{m,u}}|\mathbf{p}_{\mathrm{U}}) $ is represented as
\begin{equation}
	\label{geometric_factornode}
p(\theta_{{m,u}}|\mathbf{p}_{\mathrm{U}})=\delta\left(\theta_{{m,u}}-\frac{(\mathbf{p}_{\mathrm{U}}-\mathbf{p}_{\mathrm{BS},m})^\mathrm{T}\mathbf{e}_{u}}{\left\|\mathbf{p}_{\mathrm{U}}-\mathbf{p}_{\mathrm{BS},m}\right\|}\right), u\in \{\mathrm{x},\mathrm{y}\}.
\end{equation}
 The complex channel gain ${\alpha_{m}}$ is assigned with a complex Gaussian prior, i.e., $p(\alpha_{m})=\mathcal{CN}(0,\sigma_{{\alpha}}^2)$, $\forall m \in \mathcal{I}_M$. The UE location is assigned with a non-informative Gaussian prior with zero mean and a relatively large variance $\sigma_{\mathrm{U}}^{2}$, i.e., $p(\mathbf{p}_{\mathrm{U}})=\mathcal{N}(0,\sigma_{\mathrm{U}}^{2}\mathbf{I}_{3})$.
\par Based on the above probability models, the joint pdf is given by
\begin{multline}\label{pdf}
p\left( \mathbf{p}_{\mathrm{U}},\boldsymbol{y},\boldsymbol{\theta },\boldsymbol{\alpha} \right) =
\prod_{m=1}^{M} p\left( \boldsymbol{y}_{m,\mathrm{F}}\mid \theta_{m,\mathrm{x}},\theta_{m,\mathrm{y}},\alpha_{m}  \right)p(\alpha_{m}) \\
\times {{p(\theta_{m,\mathrm{x}}|\mathbf{p}_{\mathrm{U}})p(\theta_{m,\mathrm{y}}|\mathbf{p}_{\mathrm{U}})}}p(\mathbf{p}_{\mathrm{U}}),       
\end{multline}
where $\boldsymbol{\alpha} = [\alpha_1,\alpha_2,\dots,\alpha_{M}]^{\mathrm{T}}$. Following Bayes' theorem, the posterior distribution of the UE location $\mathbf{p}_{\mathrm{U}}$ is
\begin{equation}
	\label{17_n}
	p(\mathbf{p}_{\mathrm{U}} | \boldsymbol{y}) = \int\frac{p\left( \mathbf{p}_{\mathrm{U}},\boldsymbol{y},\boldsymbol{\theta },\boldsymbol{\alpha} \right)}{p(\boldsymbol{y})}\mathrm{d}\boldsymbol{\theta}\mathrm{d}\boldsymbol{\alpha}.
\end{equation}
An estimate of $\mathbf{p}_{\mathrm{U}}$ can be obtained by using either the minimum mean-square error (MMSE) or maximum \textit{a posteriori} (MAP) principles. However, exact posterior estimation is computationally intractable due to the high-dimensional integral. To address this issue, we adopt a low-complexity approach based on message passing, as detailed in the next section.

\section{APLE Algorithm for Subarray {Far-Field} Model}\label{S4}
\par
\begin{figure}[t]
    \centering    
    \includegraphics[width=.93\linewidth]{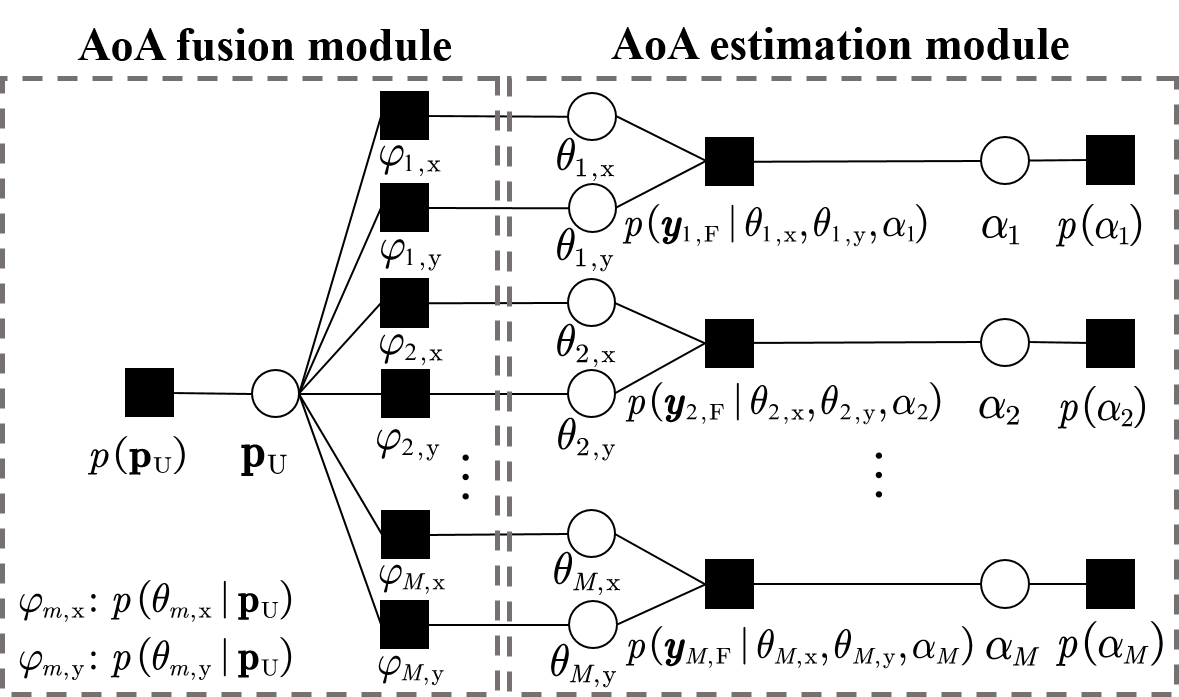}
    \caption{Factor graph representation of \eqref{pdf}.}
    \label{factor graph}
\end{figure}

 In this section, we propose the message-passing-based APLE algorithm. The factor graph representation of \eqref{pdf} is illustrated in Fig. \ref{factor graph}, where circles and squares represent variable nodes and factor nodes, respectively. The APLE algorithm is derived by performing the sum-product rule \cite{sumproduct} on the factor graph. The factor graph consists of two modules, i.e., the AoA estimation module, which aims to derive the AoA estimates at various subarrays from the received signal, and the AoA fusion module, which determines the UE location by fusing the AoA estimates based on the geometric constraints between the UE and the subarrays. For simplicity, we represent the factor node $p\left(\theta_{{m,u}}|\mathbf{p}_{\mathrm{U}}\right)$ by $\varphi_{m,u}$, for $u\in \{\mathrm{x},\mathrm{y}\}$. Denote by $\Delta_{a \rightarrow b}\left(\cdot\right)$ the message of the variable $b$ from node $a$ to $b$. The mean vector and the covariance matrix of message $\Delta_{a \rightarrow b}\left(\cdot\right)$ are denoted by $\mathbf{m}_{{a \rightarrow b}}$ and $\mathbf{C}_{{a \rightarrow b}}$, respectively.

\subsection{AoA Estimation Module}
\par We first consider the message $\Delta_{\theta_{m,\mathrm{x}}\rightarrow\varphi_{m,\mathrm{x}}}$ from the variable node $\theta_{m,\mathrm{x}}$ to the factor node $\varphi_{m,\mathrm{x}}$.
Following the sum-product rule, we have
\begin{align}  
\Delta_{\theta_{m,\mathrm{x}}\rightarrow\varphi_{m,\mathrm{x}}} \propto\int_{\theta_{m ,\mathrm{y}}}\int_{\alpha_{m}} &p(\boldsymbol{y}_{m,\mathrm{F}}|\theta_{m,\mathrm{x}},\theta_{m ,\mathrm{y}},\alpha_{m}) \notag\\
&\times p(\alpha_{m}){\Delta _{\varphi_{m ,\mathrm{y}}\rightarrow \theta_{m ,\mathrm{y}}}}. \label{1}     
\end{align}  
The integral on the right-hand side (RHS) of \eqref{1} has no closed-form expression. To facilitate subsequent message passing, we propose an approximate method for computing $\Delta_{\theta_{m,\mathrm{x}}\rightarrow\varphi_{m,\mathrm{x}}}$. Specifically, by treating the message $\Delta _{\varphi_{m ,\mathrm{y}}\rightarrow \theta_{m ,\mathrm{y}}}$ as an estimate of the prior distribution of $\theta_{m ,\mathrm{y}}$, the RHS of \eqref{1} can be viewed as an estimate of $p(\boldsymbol{y}_{m,\mathrm{F}}|\theta_{m,\mathrm{x}})$. By treating the message $\Delta _{\varphi_{m ,\mathrm{x}}\rightarrow \theta_{m ,\mathrm{x}}}$ as an estimate of the prior distribution of $\theta_{m ,\mathrm{x}}$ similarly, $\Delta_{\theta_{m,\mathrm{x}}\rightarrow\varphi_{m,\mathrm{x}}}$ can be further expressed as
\begin{subequations}\label{post}
    \begin{align}   \Delta_{\theta_{m,\mathrm{x}}\rightarrow\varphi_{m,\mathrm{x}}}&\propto \frac{\hat{p}(\boldsymbol{y}_{m,\mathrm{F}}|\theta_{m,\mathrm{x}})\Delta _{\varphi_{m ,\mathrm{x}}\rightarrow \theta_{m ,\mathrm{x}}}}{\Delta _{\varphi_{m,\mathrm{x}}\rightarrow \theta_{m,\mathrm{x}}}},\label{13_1} \\
&\propto \frac{ \hat{p}(\theta_{m,\mathrm{x}}|\boldsymbol{y}_{m,\mathrm{F}}) }{\Delta _{\varphi_{m,\mathrm{x}}\rightarrow \theta_{m,\mathrm{x}}}},\label{13_2}   
    \end{align}
\end{subequations}
where $\hat{p}(\theta_{m,\mathrm{x}}|\boldsymbol{y}_{m,\mathrm{F}})$ is an approximated posterior distribution of $\theta_{m,\mathrm{x}}$. The computation of $\hat{p}(\theta_{m,\mathrm{x}}|\boldsymbol{y}_{m,\mathrm{F}})$ in \eqref{13_2} can be taken as a Bayesian line spectra estimation problem. The multidimensional variational line spectra estimation (MVALSE) algorithm proposed in \cite{MVALSE} is employed to obtain $\hat{p}(\theta_{m,\mathrm{x}}|\boldsymbol{y}_{m,\mathrm{F}})$, which is a VM distribution given by
\begin{subequations}\label{aposterior}
  \begin{align}	
\hat{p}(\theta_{m,\mathrm{x}}|\boldsymbol{y}_{m,\mathrm{F}}) &= \mathcal{M}\left(\pi \theta_{m,\mathrm{x}};{\mu} _{\theta_{m,\mathrm{x}}},\kappa _{\theta_{m,\mathrm{x}}} \right) \\ 
 &= \frac{\exp(\kappa_{\theta_{m,\mathrm{x}}}\!\cos(\pi\theta_{m ,\mathrm{x}}\!-\!{\mu} _{\theta_{m,\mathrm{x}}}))}{2\pi I_0 (\kappa_{\theta_{m,\mathrm{x}}})}. 
\end{align}  
\end{subequations}
In \eqref{aposterior}, $\theta_{m,\mathrm{x}}$ is scaled by constant $\pi$ to meet the standard expression of a VM distribution; $I_0$, ${\mu} _{ \theta_{m,\mathrm{x}}}$, and $\kappa_{\theta_{m,\mathrm{x}}}$ represent the modified Bessel function of the first kind and order $0$, the mean direction parameter, and the concentration parameter, respectively. 
For the denominator of \eqref{13_2}, the message $\Delta _{\varphi_{m,\mathrm{x}}\rightarrow \theta_{m,\mathrm{x}}}$ (as shown later in \eqref{119}-\eqref{23}) is also a VM distribution, i.e. 
\begin{equation}\label{VM}
 \Delta _{\varphi_{m,\mathrm{x}}\rightarrow \theta_{m,\mathrm{x}}}\propto \mathcal{M}\left(\pi \theta_{m,\mathrm{x}};{\mu} _{\varphi_{m ,\mathrm{x}}\rightarrow \theta_{m ,\mathrm{x}}},\kappa _{\varphi_{m ,\mathrm{x}}\rightarrow \theta_{m ,\mathrm{x}}} \right).   
\end{equation} 
Then, based on \eqref{post} and the closure of the VM distribution under multiplication\footnotemark, we have
\footnotetext{The product of two VM pdfs is proportional to another VM pdf, i.e., $\mathcal{M}(\theta;\mu_1,\kappa_1)\mathcal{M}(\theta;\mu_2,\kappa_2) \propto \mathcal{M}(\theta;\mu_3,\kappa_3)$, where $\kappa_3 = |\kappa_1 e^{\jmath \mu_1} + \kappa_2 e^{\jmath \mu_2}|$ and $\mu_3 = \angle(\kappa_1 e^{\jmath \mu_1} + \kappa_2 e^{\jmath \mu_2})$.}
\begin{subequations}\label{16}
 \begin{align}
\Delta_{\theta_{m,\mathrm{x}}\rightarrow\varphi_{m,\mathrm{x}}} & \propto \!\frac{\mathcal{M}\!\left(\pi \theta_{m,\mathrm{x}};{\mu} _{\theta_{m ,\mathrm{x}}},\kappa _{\theta_{m ,\mathrm{x}}} \right)}{\mathcal{M}\!\left(\pi \theta_{m,\mathrm{x}};{\mu} _{\varphi_{m ,\mathrm{x}}\rightarrow \theta_{m ,\mathrm{x}}},\kappa _{\varphi_{m ,\mathrm{x}}\rightarrow \theta_{m ,\mathrm{x}}} \right)}\label{16_1} \\
 &\propto{\mathcal{M} \!\left(\pi \theta_{m,\mathrm{x}};{\mu} _{\theta_{m,\mathrm{x}}\rightarrow\varphi_{m,\mathrm{x}}},\kappa _{\theta_{m,\mathrm{x}}\rightarrow\varphi_{m,\mathrm{x}}} \right)},\label{16_2} 
\end{align}   
\end{subequations}
where ${\mu} _{\theta_{m,\mathrm{x}}\rightarrow\varphi_{m,\mathrm{x}}}$ and $\kappa _{\theta_{m,\mathrm{x}}\rightarrow\varphi_{m,\mathrm{x}}}$ satisfy 
\begin{align}\label{17}
&\kappa_{\theta_{m,\mathrm{x}}\rightarrow\varphi_{m,\mathrm{x}}}e^{\jmath{\mu} _{\theta_{m,\mathrm{x}}\rightarrow\varphi_{m,\mathrm{x}}}}\notag\\
&= \kappa_{\theta_{m ,\mathrm{x}}}e^{\jmath{\mu} _{\theta_{m ,\mathrm{x}}}} - \kappa_{\varphi_{m ,\mathrm{x}}\rightarrow \theta_{m ,\mathrm{x}}}e^{\jmath{\mu}_{\varphi_{m ,\mathrm{x}}\rightarrow \theta_{m ,\mathrm{x}}}}.    
\end{align}
\par The message $\Delta_{\theta_{m,\mathrm{y}}\rightarrow\varphi_{m,\mathrm{y}}}$ can be obtained by replacing the subscript $\mathrm{x}$ with the subscript $\mathrm{y}$ in \eqref{16} and \eqref{17}. Furthermore, the MVALSE algorithm can also be used to estimate the channel coefficient $\alpha_m$.

\subsection{AoA Fusion Module }
Given $p(\theta_{{m,u}}|\mathbf{p}_{\mathrm{U}})$ in \eqref{geometric_factornode} and $\Delta_{\theta_{{m,u}}\rightarrow\varphi_{m,u}}$ in \eqref{16_2}, the message from $\varphi_{m,u}$ to $\mathbf{p}_{\mathrm{U}}$ can be expressed as
\begin{subequations}
\begin{align}
\hspace{-10cm}
&\Delta _{\varphi_{m,u}\rightarrow \mathbf{p}_{\mathrm{U}}}\notag \\
&\propto \int_{\theta_{{m,u}}}{p(\theta_{{m,u}}|\mathbf{p}_{\mathrm{U}})\Delta_{\theta_{{m,u}}\rightarrow\varphi_{m,u}}} \label{18_1}\\
&\propto{\mathcal{M}\!\left(\!\frac{\pi(\mathbf{p}_{\mathrm{U}}-\mathbf{p}_{\mathrm{BS},m})^\mathrm{T}\mathbf{e}_{u}}{\left\|\mathbf{p}_{\mathrm{U}}-\mathbf{p}_{\mathrm{BS},m}\right\|};{\mu} _{\theta_{{m,u}}\rightarrow\varphi_{m,u}},\kappa _{\theta_{{m,u}}\rightarrow\varphi_{m,u}}\! \right)}\label{18_2}.
\end{align}    
\end{subequations}
Define $\mathcal{A} \triangleq\{(m', u') | m'\in \mathcal{I}_{M}, u' \in\{\mathrm{x}, \mathrm{y}\}\}$. Then, the message from  $\mathbf{p}_{\mathrm{U}}$ to $\varphi_{m,u}$ is computed as
\begin{subequations}
\begin{align}
\hspace{-10cm}
	&\Delta _{\mathbf{p}_{\mathrm{U}}\rightarrow\varphi_{m,u} }\notag \\
 &\propto p(\mathbf{p}_{\mathrm{U}}) \prod_{(n,v)\in\mathcal{A}\setminus(m,u)}{\Delta_{\varphi_{n,v} \rightarrow \mathbf{p}_{\mathrm{U}}}}\label{17_1}\\
 &\propto \prod_{(n,v)\in\mathcal{A}\setminus(m,u)}{\Delta_{\varphi_{n,v} \rightarrow \mathbf{p}_{\mathrm{U}}}}\label{17_2}\\
 &\propto\exp{\!\left\{\!\sum_{(n,v)\in\mathcal{A}\setminus(m,u)}{\!\!\!\!\! \kappa _{\theta_{n,v}\rightarrow \varphi_{n,v}}\!\cos(\pi\theta_{n,v}\!\!-\!\!{\mu} _{\theta_{n,v}\rightarrow \varphi_{n,v}})}\!\right\}}\label{17_3},
\end{align}    
\end{subequations}
where \eqref{17_2} holds by considering the non-informative Gaussian prior  $p(\mathbf{p}_{\mathrm{U}})$, and \eqref{17_3} is from \eqref{18_2} with $\theta_{n,v} = \frac{(\mathbf{p}_{\mathrm{U}}-\mathbf{p}_{\mathrm{BS},n})^\mathrm{T}\mathbf{e}_v}{\left\|\mathbf{p}_{\mathrm{U}}-\mathbf{p}_{\mathrm{BS},n}\right\|}$.
To simplify subsequent message updates, we approximate $\Delta _{\mathbf{p}_{\mathrm{U}}\rightarrow\varphi_{m,u} }$ by the following Gaussian pdf 
\begin{equation}\label{15}
 \Delta _{\mathbf{p}_{\mathrm{U}}\rightarrow\varphi_{m,u} }\propto\mathcal{N}(\mathbf{p}_{\mathrm{U}};\mathbf{m}_{\mathbf{p}_{\mathrm{U}}\rightarrow\varphi_{m,u}},\mathbf{C}_{\mathbf{p}_{\mathrm{U}}\rightarrow\varphi_{m,u}}).  
\end{equation}
Note that the mean $\mathbf{m}_{\mathbf{p}_{\mathrm{U}}\rightarrow\varphi_{m,u}}$ and covariance $\mathbf{C}_{\mathbf{p}_{\mathrm{U}}\rightarrow\varphi_{m,u}}$ cannot be obtained by using moment matching since it is intractable to compute the first and second moments of \eqref{17_3}. Thus, we take the following method to obtain $\mathbf{m}_{\mathbf{p}_{\mathrm{U}}\rightarrow\varphi_{m,u}}$ and $\mathbf{C}_{\mathbf{p}_{\mathrm{U}}\rightarrow\varphi_{m,u}}$. Specifically, we use the GA method to solve the following optimization problem$\colon$
\begin{equation}\label{opt}
  \hat{\mathbf{p}}_{\mathrm{U},(m,u)} =  \arg \max _{\mathbf{p}_{\mathrm{U}}} \  \varpi_{m,u}(\mathbf{p}_{\mathrm{U}}),
\end{equation}
where $\varpi_{m,u}(\cdot)$ denotes the exponential function at the RHS of \eqref{17_3}, and $\hat{\mathbf{p}}_{\mathrm{U},(m,u)}$ is the local maxima of $\varpi_{m,u}(\mathbf{p}_{\mathrm{U}})$ given by GA. 
Then, $\mathbf{m}_{\mathbf{p}_{\mathrm{U}}\rightarrow\varphi_{m,u}}$ and $\mathbf{C}_{\mathbf{p}_{\mathrm{U}}\rightarrow\varphi_{m,u}}$ are approximated respectively as
    \begin{align}
\mathbf{m}_{\mathbf{p}_{\mathrm{U}}\rightarrow\varphi_{m,u}} &=  \hat{\mathbf{p}}_{\mathrm{U},(m,u)} ~ \textrm{and} \label{20_1}\\
\mathbf{C}_{\mathbf{p}_{\mathrm{U}}\rightarrow\varphi_{m,u}} &= \left(\left.-\mathbf{H}(\mathbf{p}_{\mathrm{U}})\right|_{\mathbf{p}_{\mathrm{U}}=\hat{\mathbf{p}}_{\mathrm{U},(m,u)}}
\right)^{-1}, \label{20_2}     
    \end{align}
where $\mathbf{H}(\mathbf{p}_{\mathrm{U}})$ is the Hessian matrix of $\varpi_{m,u}(\mathbf{p}_{\mathrm{U}})$. The derivations of \eqref{20_1} and \eqref{20_2} are provided in Appendix \ref{appA}.
\par We now consider the message $\Delta_{\varphi_{m,u}\rightarrow \theta_{m,u}}$ from $\varphi_{m,u}$ to $\theta_{m,u}$, given by
\begin{align}
\Delta_{\varphi_{m,u}\rightarrow \theta_{m,u}}\propto \int_{\mathbf{p}_{\mathrm{U}}}{p(\theta_{{m,u}}|\mathbf{p}_{\mathrm{U}})\Delta _{\mathbf{p}_{\mathrm{U}}\rightarrow\varphi_{m,u} }}\label{119}.
\end{align}
The integral in \eqref{119} has no closed-form expression. We employ a similar approach to simplify the calculation of $\Delta_{\varphi_{m,u}\rightarrow \theta_{m,u}}$ as used in \cite[eq. (39)]{Teng9772371}. Let $\bar{\mathbf{u}}_{m,u}=\mathbf{m}_{\mathbf{p}_{\mathrm{U}}\rightarrow \varphi_{m,u}}-\mathbf{p}_{\mathrm{BS},m}$. The unit vector perpendicular to $\bar{\mathbf{u}}_{m,u}$ within the plane spanned by $\bar{\mathbf{u}}_{m,u}$ and $\mathbf{e}_{u}$ is denoted by $\mathbf{v}_{m,u} = \frac{\left( \bar{\mathbf{u}}_{m,u}\times \mathbf{e}_{u} \right)\times\bar{\mathbf{u}}_{m,u}}{\left\|\left( \bar{\mathbf{u}}_{m,u}\times \mathbf{e}_{u} \right)\times \bar{\mathbf{u}}_{m,u}\right\|}$, where $\times$ represents the cross product here. We focus solely on the impact of the projection of the UE localization error on $\mathbf{v}_{m,u}$. The projection is modeled by a random variable $w_{m,u}$ with $w_{m,u}=\mathbf{v}_{m,u}^\mathrm{T}(\mathbf{p}_{\mathrm{U}}-\mathbf{m}_{\mathbf{p}_{\mathrm{U}}\rightarrow \varphi_{m,u}})$. From \eqref{15}, we have $p(w_{m,u})=\mathcal{N}(0,\mathbf{v}_{m,u}^\mathrm{T}\mathbf{C}_{\mathbf{p}_{\mathrm{U}}\rightarrow \varphi_{m,u}}\mathbf{v}_{m,u})$. Given the mean AoA $\bar{\theta}_{m,u} = \frac{(\mathbf{m}_{\mathbf{p}_{\mathrm{U}}\rightarrow \varphi_{m,u}}-\mathbf{p}_{\mathrm{BS},m})^\mathrm{T}\mathbf{e}_u}{\left\|\mathbf{m}_{\mathbf{p}_{\mathrm{U}}\rightarrow \varphi_{m,u}}-\mathbf{p}_{\mathrm{BS},m}\right\|}$, the geometric constraint between $\theta_{{m,u}}$ and $w_{m,u}$ is denoted by $p(\theta_{{m,u}}|w_{m,u})=\delta\left(w_{m,u} - \left\|\bar{\mathbf{u}}_{m,u}\right\|\tan(\arccos{\bar{\theta}_{m,u}}-\arccos{\theta _{m,u}})\right)$ for a sufficiently large $\left\|\bar{\mathbf{u}}_{m,u}\right\|$ under the SFA. Then,  $\Delta_{\varphi_{m,u}\rightarrow \theta_{m,u}}$ is reduced to 
\begin{subequations}
\begin{align}
&\Delta_{\varphi_{m,u}\rightarrow \theta_{m,u}}\notag \\
&\propto \int_{w_{m,u}}{p(\theta_{{m,u}}|w_{m,u}) p(w_{m,u})}\label{22_1}\\
&\propto \exp\! \left(\!-\frac{\left\|\bar{\mathbf{u}}_{m,u}\right\|^2\tan^2(\arccos{\bar{\theta}} _{m,u}\!-\!\arccos{\theta _{m,u}})}{2\mathbf{v}_{m,u}^\mathrm{T}\mathbf{C}_{\mathbf{p}_{\mathrm{U}}\rightarrow \varphi_{m,u}}\mathbf{v}_{m,u}}\!\right)\label{22_2}\\
&\propto\mathcal{M}\left(\pi\theta_{m,u};{\mu}_{\varphi_{m,u}\rightarrow \theta_{m,u}},\kappa_{\varphi_{m,u}\rightarrow \theta_{m,u}}\right)\label{22_3},   
\end{align}
\end{subequations}
where \eqref{22_3} holds by approximating \eqref{22_2} as a VM distribution. In \eqref{22_3},
\begin{subequations}\label{23}
    \begin{align}
     {\mu} _{\varphi_{m,u}\rightarrow \theta_{m,u}}&=  \pi\bar{\theta} _{m,u},\label{23_1} \\    \kappa_{\varphi_{m,u}\rightarrow\theta_{m,u}} &=\frac{\left\|\bar{\mathbf{u}}_{m,u}\right\|^2}{\pi^2\left(1-\bar{\theta} _{m,u}^2\right)\mathbf{v}_{m,u}^\mathrm{T}\mathbf{C}_{\mathbf{p}_{\mathrm{U}}\rightarrow\varphi_{m,u}}\mathbf{v}_{m,u}} \label{23_2},
    \end{align}
\end{subequations}
where $\bar{\theta} _{m,u}$ is the maximum point of \eqref{22_2}, \eqref{23_2} follows by equating \eqref{22_2}, and \eqref{22_3} based on Taylor series expansion at $\bar{\theta}_{m,u}$.
\addtolength{\topmargin}{0.05in}
\begin{algorithm}[t]
	\caption{APLE Algorithm} 
	\label{APLE-Algorithm} 
	{\bf Input:} $T_1$, $T_2$, $\sigma^2$, $f$, $d_{\mathrm{x}}$, $d_{\mathrm{y}}$, $ N_{\mathrm{x}}$, $N_{\mathrm{y}}$, $M$, $\mathbf{p}_{\mathrm{BS},m}$, and $\boldsymbol{y}$.
	\\
	{\bf Output:} The UE location estimation.
	\par \begin{algorithmic}[1] 
		\REPEAT
            \STATE {\% AoA estimation module}
             \FOR {$m = 1$ to $M$}        
            \STATE { $\forall{u\in \{\mathrm{ x},\mathrm{y}\}}$, update $\hat{p}(\theta_{m,u}|\boldsymbol{y}_{m,\mathrm{F}})$ and $\hat{\alpha}_m$ by the MVALSE algorithm}.
             \STATE { $\forall{u\in \{\mathrm{ x},\mathrm{y}\}}$, update $\Delta _{\theta_{m,u}\rightarrow \varphi_{m,u}}(\theta_{m,u})$ by following \eqref{16_2}, \eqref{17}.}
            \ENDFOR
		\STATE {\% AoA fusion module}
        \FOR {$m = 1$ to $M$} 
        \STATE { $\forall{u\in \{\mathrm{ x},\mathrm{y}\}}$, update $\Delta _{\varphi_{m,u}\rightarrow \mathbf{p}_{\mathrm{U}}}$ by \eqref{18_2}.}
		\STATE { $\forall{u\in \{\mathrm{ x},\mathrm{y}\}}$, update $\hat{\mathbf{p}}_{\mathrm{U},(m,u)}$ by solving the optimization problem in \eqref{opt} via GA.}      
		\STATE { $\forall{u\in \{\mathrm{ x},\mathrm{y}\}}$, update $\Delta _{\mathbf{p}_{\mathrm{U}}\rightarrow\varphi_{m,u} }$ by \eqref{20_1}, \eqref{20_2}.}
		\STATE { $\forall{u\in \{\mathrm{ x},\mathrm{y}\}}$, update $\Delta _{\varphi_{m,u}\rightarrow \theta_{m,u}}$ by \eqref{22_3},\eqref{23_1},\eqref{23_2}.}
        \ENDFOR
        \UNTIL the maximum number of iterations $T_1$ is reached.
		\STATE \textbf{return}
	\end{algorithmic}
\end{algorithm}
\subsection{Overall Algorithm}
The APLE algorithm is summarized in Algorithm \ref{APLE-Algorithm}. The messages are iteratively passed between the AoA estimation and AoA fusion modules until the maximum number of iterations $T_1$ is reached. The complexity of the APLE algorithm primarily arises from the AoA estimation module. The complexity of calculating $\hat{p}(\theta_{m,\mathrm{x}}|\boldsymbol{y}_{m,\mathrm{F}})$ and $\hat{p}(\theta_{m,\mathrm{y}}|\boldsymbol{y}_{m,\mathrm{F}})$ is $\mathcal{O}(T_{2}N_{m,\mathrm{x}}N_{m,\mathrm{y}})$, where $T_{2}$ is the number of iterations in MVALSE. Therefore, the overall complexity of APLE is $\mathcal{O}(T_{1}T_{2}N_{\mathrm{B}})$ with $N_{\mathrm{B}}=MN_{m,\mathrm{x}}N_{m,\mathrm{y}}$.

\section{Ehanced APLE Algorithm}\label{S5}
\subsection{Motivations}
The performance of the APLE algorithm can be further improved in two aspects.
Firstly, APLE relies on the SFM in \eqref{far}, which adopts the far-field plane-wave assumption for each subarray. This approximation can lead to performance degradation in the location estimation. Secondly, in addition to estimating the UE location, the APLE algorithm also estimates the nuisance subarray channel coefficients $\alpha_1,\alpha_2,\dots,\alpha_{M}$, by treating them as independent unknown variables. As the number of subarrays $M$ increases, the number of unknowns to be estimated also increases, which may compromise the estimation performance. To address these two issues, we next propose the E-APLE algorithm. In the E-APLE algorithm, we adopt the near-field received signal model of the entire BS array as described in \eqref{ENFM}, instead of relying on the SFM to mitigate performance degradation. We formulate the ML estimation problem for the UE location based on this near-field received signal model. Subsequently, we propose a BCA method to solve the ML problem. The proposed BCA method is initialized by the location estimate obtained from APLE. The details of the E-APLE algorithm are discussed as follows.
\subsection{Problem Formulation}
In Section \ref{S2}, we discuss the log-likelihood function associated with the ML problem for different array sizes. From \eqref{llf}, the ML estimate of the UE location and the complex channel gain $\alpha$ are given by
\begin{equation}\label{MLproblem}
    \left[\hat{\mathbf{p}}_{\mathrm{U}}, \hat{{\alpha}} \right]=\arg\max _{\mathbf{p}_{\mathrm{U}}, {\alpha}}\; -\frac{1}{\sigma^2}  \vert|\boldsymbol{y}- \alpha \mathbf{a}(\mathbf{p}_{\mathrm{U}}) \vert|^2. 
\end{equation}
For a given UE location $\mathbf{p}_{\mathrm{U}}$, the optimal $\alpha$ that maximizes the objective function in \eqref{MLproblem} is given by 
\begin{equation}\label{alphaLS}
    \hat{\alpha} = \frac{\mathbf{a}^{\mathrm{H}}(\mathbf{p}_{\mathrm{U}})\boldsymbol{y}}{ \vert|\mathbf{a}(\mathbf{p}_{\mathrm{U}})\vert|^2}.
\end{equation}
By plugging \eqref{alphaLS} into \eqref{MLproblem}, the problem described in \eqref{MLproblem} is expressed as
\begin{equation}\label{FinalNLS}
       \max _{\mathbf{p}_{\mathrm{U}}} \; -\frac{1}{\sigma^2}  \Bigg\Vert\boldsymbol{y}  - \frac{\mathbf{a}^{\mathrm{H}}(\mathbf{p}_{\mathrm{U}})\boldsymbol{y}}{ \vert|\mathbf{a}(\mathbf{p}_{\mathrm{U}})\vert|^2}\mathbf{a}(\mathbf{p}_{\mathrm{U}}) \Bigg\Vert^2. 
\end{equation}
Problem \eqref{FinalNLS} generally has no closed-form solutions. Next, we adopt a BCA approach for solving problem \eqref{FinalNLS} in the distance-angle polar domain.
\subsection{Algorithm Design}
As shown in Fig. \ref{FigMLE}, the objective function of problem \eqref{FinalNLS} exhibits a special \textit{ridge} structure in the degraded two-dimensional case, where $N_{\mathrm{x}}=N_{\mathrm{B}}$ and $N_{\mathrm{y}}=1$. Specifically, the objective function is highly multi-modal in the Cartesian domain, and the global maxima occur at the true UE location. Near the true UE location, the objective function value decreases in all directions, with a particularly slow decrease along the radial direction connecting the true UE location and the origin. This ridge structure characterized by the different descending speeds is significant especially when the BS size is small. This implies that the objective function is sensitive to the UE direction, and inspires us to solve problem \eqref{FinalNLS} in the distance-angle polar domain. 

The UE location ${\mathbf{p}}_{\mathrm{U}}$ can be rewritten as ${\mathbf{p}}_{\mathrm{U}}=r\left[\cos \omega \sin \phi, \sin \omega \sin \phi, \cos\phi \right]$ in the spherical coordinate system, where $r= \|{\mathbf{p}}_{\mathrm{U}}\|$ is the distance between the UE and the origin; $\omega$ and $\phi$ denote the azimuth and polar angles of the UE, respectively, as illustrated in Fig. \ref{system}. Then, an equivalent problem formulation of \eqref{FinalNLS} in the distance-angle polar domain is given by 
\begin{subequations}\label{PolarNLS}
    \begin{align}
        \max_{r, \boldsymbol{\vartheta}} &~  \mathcal{F}({r}, \boldsymbol{\vartheta}) \\
        \text { s. t. } & r>0, ~\omega\in[0,2\pi),~ \phi\in[0,\pi/2),
    \end{align}
\end{subequations}
 where $\boldsymbol{\vartheta}\triangleq [\omega,\phi]^{\mathrm{T}}$, and
 \begin{equation}
 \mathcal{F}({r},  \boldsymbol{\vartheta}) \triangleq
     -\frac{1}{\sigma^2}  \Bigg\Vert\boldsymbol{y}  - \frac{\mathbf{a}^{\mathrm{H}}(r,  \boldsymbol{\vartheta})\boldsymbol{y}}{ \vert|\mathbf{a}(r,  \boldsymbol{\vartheta})\vert|^2}\mathbf{a}(r,  \boldsymbol{\vartheta}) \Bigg\Vert^2.
 \end{equation}
In the distance-angle domain, the objective function $\mathcal{F}({r}, \boldsymbol{\vartheta})$ has relatively independent modes for the distance $r$ and the angle $\boldsymbol{\vartheta}$, i.e., the optimal value of $\boldsymbol{\vartheta}$ that maximizes $\mathcal{F}({r}, \boldsymbol{\vartheta})$ tends to be independent of the value of $r$. This means that searching $\mathcal{F}({r}, \boldsymbol{\vartheta})$ along the distance and angle coordinates separately, rather than jointly, may more efficiently find its maxima. Inspired by this, we propose to take a BCA approach for solving problem \eqref{PolarNLS}. In particular, we divide problem \eqref{PolarNLS} into two sub-problems: sub-problem 1 is to optimize $ \boldsymbol{\vartheta}$ for fixed $r$, and sub-problem 2 is to optimize $r$ with $ \boldsymbol{\vartheta}$ fixed. These two sub-problems are alternately optimized until convergence. Besides, the highly multi-modal characteristic of the objective function is inherited in the distance-angle domain. To prevent the BCA method from becoming trapped in poor local maxima, we initialize the algorithm by the UE location estimate obtained from APLE. This initialization helps the algorithm begin its search from a more promising starting point. The two sub-problems and their solutions are illustrated as follows.

\subsubsection{Sub-problem 1}
For a fixed distance $\check{r}$, the sub-problem of optimizing $ \boldsymbol{\vartheta}$ is given by
\begin{subequations}\label{SP1}
    \begin{align}
        \max_{\boldsymbol{\vartheta}} &~  \mathcal{F}(\check{r},  \boldsymbol{\vartheta}) \\
        \text { s. t. } & \omega\in[0,2\pi),~ \phi\in[0,\pi/2).
    \end{align}
\end{subequations}
For problem \eqref{SP1}, we find a stationary point of $\mathcal{F}(\check{r},  \boldsymbol{\vartheta})$ with respect to $\boldsymbol{\vartheta}$ via GA. Specifically, denote by $\boldsymbol{\vartheta}^{(\mathrm{old})}$ the $\boldsymbol{\vartheta}$ obtained in the previous iteration. Denote by $\nabla \mathcal{F}_{\boldsymbol{\vartheta}^{(\mathrm{old})}} \triangleq \left[\frac {\partial \mathcal{F}(\check{r}, \boldsymbol{\vartheta})}{\partial \psi_{\mathrm{x}} }, \frac {\partial \mathcal{F}(\check{r}, \boldsymbol{\vartheta})}{\partial \psi_{\mathrm{y}} } \right]\Big|_{\boldsymbol{\vartheta}=\boldsymbol{\vartheta}^{(\mathrm{old})}}$ the gradient of $\mathcal{F}(\check{r}, \boldsymbol{\vartheta})$ with respect to $\boldsymbol{\vartheta}$. The update of $\boldsymbol{\vartheta}$ in the current iteration is given by
\begin{equation}
\boldsymbol{\vartheta}^{(\mathrm{new})}=\boldsymbol{\vartheta}^{(\mathrm{old})} + \epsilon_1 \nabla \mathcal{F}_{\boldsymbol{\vartheta}^{(\mathrm{old})}},
\end{equation}
 where $\epsilon_1>0$ is an appropriate step size that can be selected from the backtracking line search to satisfy
\begin{equation}\label{FGradient}
 \mathcal{F}\left(\check{r}, \boldsymbol{\vartheta}^{(\mathrm{new})}\right) \geq  \mathcal{F}\left(\check{r}, \boldsymbol{\vartheta}^{(\mathrm{old})}\right).
\end{equation}
An update of $\boldsymbol{\vartheta}$ is obtained after $T_{\boldsymbol{\vartheta}}$ iterations.
 \subsubsection{Sub-problem 2}
For fixed angles $\check{\boldsymbol{\vartheta}}$, the sub-problem of optimizing $r$ is given by
\begin{subequations}\label{SP2}
    \begin{align}
        \max_{r} &~  \mathcal{F}\left({r}, \check{\boldsymbol{\vartheta}}\right) \\
        \text { s. t. } & r>0.
    \end{align}
\end{subequations}
Problem \eqref{SP1} can be solved through GA in a similar manner as that used for sub-problem 1. An update of $r$ is obtained after $T_{r}$ iterations.

\begin{algorithm}[t]
	\caption{E-APLE Algorithm} 
	\label{GDA} 
	{\bf Input:}  $T_{\boldsymbol{\vartheta}}$, $T_{r}$, $T_3$, $\sigma^2$, $f$, $d_{\mathrm{x}}$, $d_{\mathrm{y}}$, $ N_{\mathrm{x}}$, $N_{\mathrm{y}}$, and $\boldsymbol{y}$.
	\\
	{\bf Output:} The estimates of $r$ and $\boldsymbol{\vartheta}$.
	\par \begin{algorithmic}[1] 
 \STATE{Obtain the location estimate $\hat{\mathbf{p}}_{\mathrm{U};\mathrm{APLE}}=[\hat{x},\hat{y},\hat{z}]^{\mathrm{T}}$ from APLE.}
    \STATE {Initialization: $\hat{r}=\|\hat{\mathbf{p}}_{\mathrm{U};\mathrm{APLE}}\|$, $\hat{\boldsymbol{\vartheta}}= \left[ \arctan\left({\hat{y}}/{\hat{x}}\right), \arccos\left( \hat{z}/\hat{r}\right) \right]^{\mathrm{T}}$.}
    \REPEAT
        \STATE {\% Update $\boldsymbol{\vartheta}$}
        \STATE {Fix $\check{r}=\hat{r}$ and compute $\nabla \mathcal{F}_{\boldsymbol{\vartheta}}$.}
        \REPEAT
        \STATE {Find $\epsilon_1>0$ that satisfies \eqref{FGradient} using the backtracking line search.}
        \STATE {Update $\hat{\boldsymbol{\vartheta}}=\hat{\boldsymbol{\vartheta}}+\epsilon_1 \nabla \mathcal{F}_{\boldsymbol{\vartheta}}$}.
        \UNTIL maximum number of iterations $T_{\boldsymbol{\vartheta}}$ is reached.
        \STATE {\% Update $r$}
        \STATE {Fix $\check{\boldsymbol{\vartheta}}=\hat{\boldsymbol{\vartheta}}$ and compute $\nabla \mathcal{F}_{r} $.}
        \REPEAT
        \STATE {Find $\epsilon_2>0$ using the backtracking line search.}
        \STATE {Update $\hat{r}=\hat{r}+\epsilon_2 \nabla \mathcal{F}_{r}$}.
        \UNTIL maximum number of iterations $T_{r}$ is reached.
        \UNTIL the maximum number of iterations $T_3$ is reached.
		\STATE \textbf{return} $\hat{r}$, $\hat{\boldsymbol{\vartheta}}$.
	\end{algorithmic}
\end{algorithm}

The BCA-based E-APLE algorithm is summarized in Algorithm \ref{GDA}. In Line $1$, we obtain the location estimate $\hat{\mathbf{p}}_{\mathrm{U};\mathrm{APLE}}$ from APLE. $\hat{\mathbf{p}}_{\mathrm{U};\mathrm{APLE}}$ is then used to initialize the E-APLE algorithm in Line $2$. From Line $3$ to Line $16$, the E-APLE algorithm alternately updates the estimates of $r$ and $\boldsymbol{\vartheta}$ until the maximum number of iterations $T_3$ is reached. In each iteration, $r$ and $\boldsymbol{\vartheta}$ can be updated once or multiple times through GA. The final estimate of the UE location is given by $\hat{\mathbf{p}}_{\mathrm{U}}=\hat{r}\left[\cos \hat\omega \sin \hat\phi, \sin \hat\omega \sin \hat\phi, \cos\hat\phi \right]$.  

The complexity of the E-APLE algorithm primarily arises from the utilization of APLE for initialization and BCA. The complexity of obtaining the initialization of $\mathbf{p}_{\mathrm{U}}$ is  $\mathcal{O}(T_{1}T_{2}N_{\mathrm{B}})$. The complexity of BCA is  $\mathcal{O}( T_{3}(T_{\boldsymbol{\vartheta}}+T_{r})N_{\mathrm{B}} )$, where $T_{3}$ is the number of iterations of BCA. The total complexity of the E-APLE algorithm is given by $\mathcal{O}( (T_{1}T_{2} + T_{3}(T_{\boldsymbol{\vartheta}}+T_{r}) )N_{\mathrm{B}})$, which is still linear with the number of BS antennas.

\section{Lower Bounds Analysis}\label{S6}
In this section, we derive the CRB for the considered near-field localization problem, serving as a benchmark for the APLE and E-APLE algorithms. Since the APLE algorithm is developed based on the simplified SFM, we additionally establish the MCRB as a benchmark for evaluating the performance of the APLE algorithm. This is achieved by following the methodology outlined in the MCRB analysis literature \cite{MCRB, PMCRB}.

\subsection{CRB Analysis}
We first study the CRB under the near-field signal model \eqref{ENFM}. Define the parameter vector for the considered problem $\boldsymbol{\eta}=\big[\mathbf{p}_{\mathrm{U}}^{\mathrm{T}},\angle \alpha,\vert\alpha\vert\big]^{\mathrm{T}}$, where $\angle \alpha$ and $\vert\alpha\vert$ represent the angle and amplitude of the equivalent complex channel gain $\alpha$, respectively. Denote by $\bar{\boldsymbol{\eta}}=\big[\bar{\mathbf{p}}_{\mathrm{U}}^{\mathrm{T}},\angle \bar{\alpha},\vert\bar{\alpha}\vert\big]^{\mathrm{T}}$ the ground truth value of $\boldsymbol{\eta}$. The mean square error of an unbiased estimate of the parameter $\eta_{q}$ is lower bounded by the CRB, which corresponds to the $q$-th diagonal element of the inverse of the Fisher information matrix (FIM) $\mathbf{J} \in\mathbb{R}^{5\times5}$, $q\in \mathcal{I}_{5}$. Given the signal model in~\eqref{ENFM},  we have $\boldsymbol{\mu}(\boldsymbol{\eta})  =\alpha \mathbf{a}(\mathbf{p}_{\mathrm{U}})$ and $\boldsymbol{\Sigma} =\sigma^2\mathbf{I}_{N_{\mathrm{B}}}$. Since $\boldsymbol{\Sigma} $ is irrelevant to the parameter vector $\boldsymbol{\eta} $, the FIM of $\boldsymbol{\eta} $ is given by \cite{detection}
\begin{align}
	\label{FIMcal}
	\mathbf{J}   =\frac{2}{\sigma ^2} \text{Re}\left[ \left( \frac{\partial{\boldsymbol\mu}({\boldsymbol\eta})}{\partial \boldsymbol{\eta} } \right)^{\mathrm{H}} \frac{\partial{\boldsymbol\mu}({\boldsymbol\eta})}{\partial \boldsymbol{\eta} } \right].
\end{align}
Denote by $\mu_{t}$ the $t$-th element of $\boldsymbol{\mu}$ with $t=\left(\left(i-1\right) N_{\mathrm{y}}+j\right)$, $t\in \mathcal{I}_{N_{\mathrm{B}}}$, $i\in \mathcal{I}_{N_{\mathrm{x}}}$, $j\in \mathcal{I}_{N_{\mathrm{y}}}$. The derivatives of $\mu_{t}$ with respect to $\mathbf{p}_{\mathrm{U}}$, $\angle \alpha$ and $\vert\alpha\vert$ are given by
\begin{subequations}
\label{44}
\begin{align}
   \frac{\partial \mu_{t} }{\partial \mathbf{p}_{\mathrm{U}}}
    &=   -\frac{\jmath 2\pi  (\mathbf{p}_{\mathrm{U}}-\mathbf{p}_{\mathrm{BS},(i,j)})}{\lambda \left\|\mathbf{p}_{\mathrm{U}}-\mathbf{p}_{\mathrm{BS},(i,j)}\right\|} \alpha e^{-\jmath\frac{2\pi}{\lambda}\left\|\mathbf{p}_{\mathrm{U}}-\mathbf{p}_{\mathrm{BS},(i,j)}\right\|}  
    ,\\
\frac{\partial \mu_{t} }{\partial \vert\alpha\vert}&= \frac{\alpha}{\vert\alpha\vert} e^{-\jmath\frac{2\pi}{\lambda}\left\|\mathbf{p}_{\mathrm{U}}-\mathbf{p}_{\mathrm{BS},(i,j)}\right\| } ,\\
\frac{\partial \mu_{t} }{\partial \angle\alpha}&=\jmath\alpha e^{-\jmath\frac{2\pi}{\lambda}\left\|\mathbf{p}_{\mathrm{U}}-\mathbf{p}_{\mathrm{BS},(i,j)}\right\|} , 
\end{align}
\end{subequations}
respectively, $t\in \mathcal{I}_{N_{\mathrm{B}}}$, $i\in \mathcal{I}_{N_{\mathrm{x}}}$, $j\in \mathcal{I}_{N_{\mathrm{y}}}$. With the FIM $\mathbf{J}$ in \eqref{FIMcal}, the CRB of the estimate $\hat{\boldsymbol{\eta}}$ is given by
\begin{equation}\label{CRB}
	\mathbb{E}\Big[ (\hat{\boldsymbol{\eta}}-\bar{\boldsymbol{\eta}})(\hat{\boldsymbol{\eta}}-\bar{\boldsymbol{\eta}})^{\mathrm{T}}\Big] \geq \mathbf{J}^{-1},
\end{equation}
which serves as a performance lower bound of the proposed algorithms.

\subsection{MCRB for the SFM in \eqref{far}}
 We now study the MCRB for the SFM in \eqref{far}. Based on \eqref{ENFM}, the received signal at the $m$-th subarray can be expressed as
\begin{align}\label{Subnear}
    \boldsymbol{y}_{m}&= \alpha_m \mathbf{a}_{m}( \mathbf{p}_{\mathrm{U}} ) +\boldsymbol{n}_{m},
\end{align}
where $\alpha_m=\alpha e^{-\jmath\frac{ 2\pi}{\lambda}{r_m}}$ is viewed as a function of $\alpha$ and $ \mathbf{p}_{\mathrm{U}}$ here; $\mathbf{a}_{m}( \mathbf{p}_{\mathrm{U}} )\in\mathbb{C}^{N_{m}}$ represents the near-field steering vector of the $m$-th subarray, with the $\left(\left(k-1\right) N_{m,\mathrm{y}}+l\right)$-th element of $\mathbf{a}_{m}(\mathbf{p}_{\mathrm{U}})$ denoted by $e^{-\jmath\frac{ 2\pi}{\lambda} (r_{m,(k,l)} - r_m)}$, $k\in \mathcal{I}_{N_{m,\mathrm{x}}}, l \in \mathcal{I}_{N_{m,\mathrm{y}}}$. For simplicity, we reshape the received signal at the BS array into a vector as $\Tilde{\boldsymbol{y}}=[\boldsymbol{y}_{1}^{\mathrm{T}},\boldsymbol{y}_{2}^{\mathrm{T}},\dots,\boldsymbol{y}_{M}^{\mathrm{T}}]^{\mathrm{T}}$. The distribution of $\Tilde{\boldsymbol{y}}$ conditioned on $\bar{\boldsymbol{\eta}}$ is given by 
\begin{equation}\label{truepdf}
p(\Tilde{\boldsymbol{y}}|\bar{\boldsymbol{\eta}})=\mathcal{CN}\left(\Tilde{\boldsymbol{y}}; \boldsymbol{\mu}_{\mathrm{N}}(\bar{\boldsymbol{\eta}}),\sigma^2\mathbf{I}_{N_{\mathrm{B}}}\right) ,
\end{equation}
where $\boldsymbol{\mu}_{\mathrm{N}}(\boldsymbol{\eta})=\big[\boldsymbol{\mu}_{1,{\mathrm{N}}}^\mathrm{T},\boldsymbol{\mu}_{2,{\mathrm{N}}}^\mathrm{T},...,\boldsymbol{\mu}_{M,{\mathrm{N}}}^\mathrm{T}\big]^\mathrm{T}$ with $\boldsymbol{\mu}_{m,{\mathrm{N}}}=\alpha_m \mathbf{a}_{m}( \mathbf{p}_{\mathrm{U}} )$, $m\in \mathcal{I}_{M}$.
Based on the SFM in \eqref{far}, which is referred to as the misspecified model, $\alpha_m$ is regarded as an independent variable, $m\in \mathcal{I}_{M}$. In this case, we define a parameter vector consisting of all the variables to be estimated as
 \begin{equation}
\boldsymbol{\gamma}=\big[\mathbf{p}_{\mathrm{U}}^{\mathrm{T}},\angle \alpha_{1},\vert\alpha_{1}\vert,\dots,\angle\alpha_{M},\vert\alpha_{M}\vert\big]^\mathrm{T}, 
 \end{equation}
 where $\angle \alpha_{m}$ and $\vert\alpha_{m}\vert$ represent the angle and amplitude of the equivalent channel coefficient $\alpha_m$, respectively, $m\in \mathcal{I}_{M}$.
The misspecified parametric pdf of $\Tilde{\boldsymbol{y}}$ is given by 
\begin{equation}\label{mispdf}
p(\Tilde{\boldsymbol{y}} |\boldsymbol{\gamma})=\mathcal{CN}\left(\Tilde{\boldsymbol{y}}; \boldsymbol{\mu}_{\mathrm{F}}(\boldsymbol{\gamma}),\sigma^2\mathbf{I}_{N_{\mathrm{B}}}\right) ,
\end{equation}
where $\boldsymbol{\mu}_{\mathrm{F}}(\boldsymbol{\gamma})=\big[\boldsymbol{\mu}_{1,{\mathrm{F}}}^\mathrm{T},\boldsymbol{\mu}_{2,{\mathrm{F}}}^\mathrm{T},...,\boldsymbol{\mu}_{M,{\mathrm{F}}}^\mathrm{T}\big]^\mathrm{T}$ with $\boldsymbol{\mu}_{m,{\mathrm{F}} }={\alpha_{m}} \mathbf{a}_{\mathrm{F}}(\theta_{m,\mathrm{x}},\theta_{m,\mathrm{y}})$ and $\theta_{m,u}=\frac{(\mathbf{p}_{\mathrm{U}}-\mathbf{p}_{\mathrm{BS},m})^\mathrm{T}\mathbf{e}_{u}}{\left\|\mathbf{p}_{\mathrm{U}}-\mathbf{p}_{\mathrm{BS},m}\right\|}$, $m\in \mathcal{I}_{M}$, $u\in \{{\mathrm{x}},{\mathrm{y}}\}$. 

The pseudo-true parameter that minimizes the Kullback–Leibler divergence between the true pdf $p(\Tilde{\boldsymbol{y}} |\bar{\boldsymbol{\eta}})$ in \eqref{truepdf} and the misspecified parametric pdf $p(\Tilde{\boldsymbol{y}} |\boldsymbol{\gamma})$ in \eqref{mispdf} can be obtained as \cite{MCRB}
\begin{equation}\label{F-eta}
    {\boldsymbol\gamma}_{0} = \arg \min_{{\boldsymbol\gamma}} \Vert {\boldsymbol\mu_{\mathrm{N}}}( {\bar{\boldsymbol\eta}}) - {\boldsymbol\mu_{\mathrm{F}}}({\boldsymbol\gamma}) \Vert^2. 
\end{equation}
Given $\bar{\boldsymbol{\eta}}$, the ground truth value of $\boldsymbol{\gamma}$ is denoted by $\bar{\boldsymbol{\gamma}}=\big[\bar{\mathbf{p}}_{\mathrm{U}}^\mathrm{T},\angle \bar{\alpha}_{1},\vert\bar{\alpha}_{1}\vert,\dots,\angle \bar{\alpha}_{M},\vert\bar{\alpha}_{M}\vert\big]^\mathrm{T}$  with $\angle \bar{\alpha}_{m}=\angle \bar{\alpha}-\frac{2\pi r_m}{\lambda}$ and $\vert\bar{\alpha}_{m}\vert=\vert\bar{\alpha}\vert$, $m\in \mathcal{I}_{M}$. The MCRB for the SFM in \eqref{far} is given by 
\begin{multline}\label{F-LB}
       \text{MCRB}(\bar{\boldsymbol{\gamma}}, {\boldsymbol\gamma}_{0})=\mathbf{A}_{{\boldsymbol\gamma}_{0}}^{-1}\mathbf{B}_{{\boldsymbol\gamma}_{0}}\mathbf{A}_{{\boldsymbol\gamma}_{0}}^{-1} + (\bar{\boldsymbol{\gamma}}- {\boldsymbol\gamma}_{0})(\bar{\boldsymbol{\gamma}}- {\boldsymbol\gamma}_{0})^{\mathrm{T}},  
\end{multline}
where $\mathbf{A}_{{\boldsymbol\gamma}_{0}}$ and $\mathbf{B}_{{\boldsymbol\gamma}_{0}}$ are two generalizations of the FIMs under the SFM \cite{PMCRB}. 
The matrices $\mathbf{A}_{{\boldsymbol\gamma}_{0}}$ and $\mathbf{B}_{{\boldsymbol\gamma}_{0}}$ can be obtained based on the pseudo-true parameter vector ${\boldsymbol\gamma}_{0}$ as
\begin{align}
    & [\mathbf{A}_{{\boldsymbol\gamma}_{0}}]_{a,b} \notag\\
    &= \!\frac{2}{\sigma^2}\! \text{Re}\!\left.\left[\boldsymbol{\epsilon}^{\mathrm{H}}({\boldsymbol\gamma})\frac{\partial^2{\boldsymbol\mu_{\mathrm{F}}}({\boldsymbol\gamma})}{\partial [\boldsymbol{\gamma}]_a \partial [\boldsymbol{\gamma}]_b}\! -\!\left(\frac{\partial{\boldsymbol\mu_{\mathrm{F}}}({\boldsymbol\gamma})}{\partial [\boldsymbol{\gamma}]_a} \right)^{\mathrm{H}}\!
        \frac{\partial{\boldsymbol\mu_{\mathrm{F}}}({\boldsymbol\gamma})}{\partial [\boldsymbol{\gamma}]_b} \right]\right|_{\boldsymbol{\gamma}= {\boldsymbol\gamma}_{0} },    
\end{align}
\begin{align}
[\mathbf{B}_{{\boldsymbol\gamma}_{0}}]_{a,b} =\! \!&\left.\frac{4}{\sigma^4}\text{Re}\!\left[\boldsymbol{\epsilon}^{\mathrm{H}}({\boldsymbol\gamma})\frac{\partial{\boldsymbol\mu_{\mathrm{F}}}({\boldsymbol\gamma})}{\partial [\boldsymbol{\gamma}]_a} \right]\!\text{Re} \!\!\left[\boldsymbol{\epsilon}^{\mathrm{H}}({\boldsymbol\gamma})\frac{\partial{\boldsymbol\mu_{\mathrm{F}}}({\boldsymbol\gamma})}{\partial [\boldsymbol{\gamma}]_b}
\!\right]\!\right|_{{\boldsymbol\gamma} = {\boldsymbol\gamma}_{0}} \notag\\
& + \!\left. \frac{2}{\sigma^2}\text{Re}\left[  \left(\frac{\partial{\boldsymbol\mu_{\mathrm{F}}}({\boldsymbol\gamma})}{\partial [\boldsymbol{\gamma}]_a} \right)^{\mathrm{H}}\frac{\partial{\boldsymbol\mu_{\mathrm{F}}}({\boldsymbol\gamma})}{\partial [\boldsymbol{\gamma}]_b} \right]\!\right|_{{\boldsymbol\gamma} = {\boldsymbol\gamma}_{0}},\label{F-matrix_B}
\end{align}
where $\boldsymbol{\epsilon}({\boldsymbol\gamma}) \triangleq {\boldsymbol\mu_{\mathrm{N}}}( {\bar{\boldsymbol\eta}}) - {\boldsymbol\mu_{\mathrm{F}}}({\boldsymbol\gamma})$, $a,b\in \mathcal{I}_{3+2M}$. Under the SFM, the mean square error of the estimate of ${\bar{\boldsymbol\gamma}}$ is bounded by the MCRB, expressed as
\begin{equation}\label{FMCRB}
\mathbb{E}\Big[ ( \hat{\boldsymbol\gamma}-{\bar{\boldsymbol\gamma}})(\hat{\boldsymbol\gamma}-{\bar{\boldsymbol\gamma}})^{\mathrm{T}}\Big] \geq \text{MCRB}(\bar{\boldsymbol{\gamma}}, {\boldsymbol\gamma}_{0}).
\end{equation}
The derived MCRB can be employed to assess the performance of the proposed APLE algorithm. 

\begin{figure}[ht]
	\centering
	\begin{subfigure}{0.9\linewidth}
		\centering
		\includegraphics[width=1\linewidth]{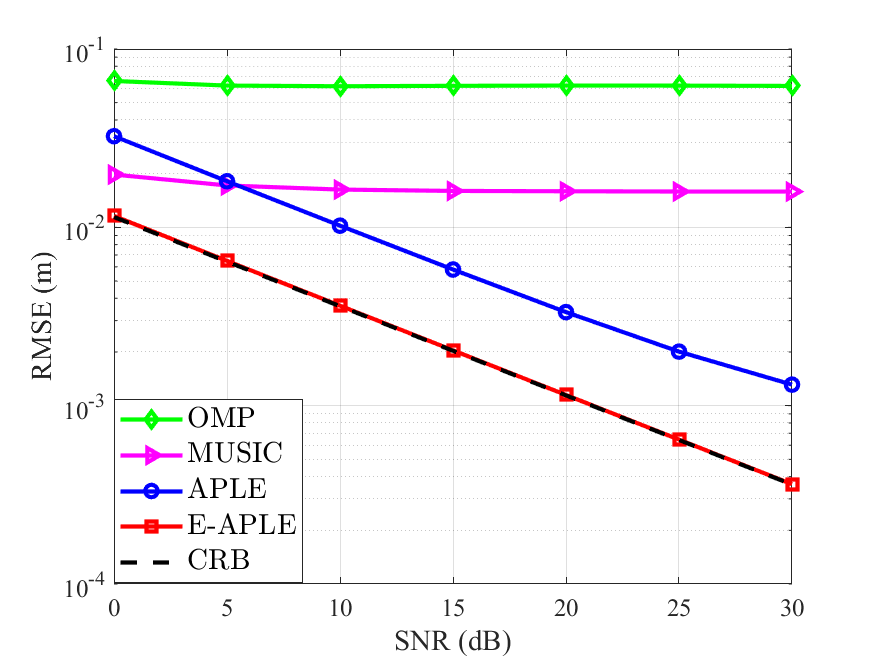}
		\caption{$r=1$ m}
  \label{1m}
	\end{subfigure}
	\centering
	\begin{subfigure}{0.9\linewidth}
		\centering
		\includegraphics[width=1\linewidth]{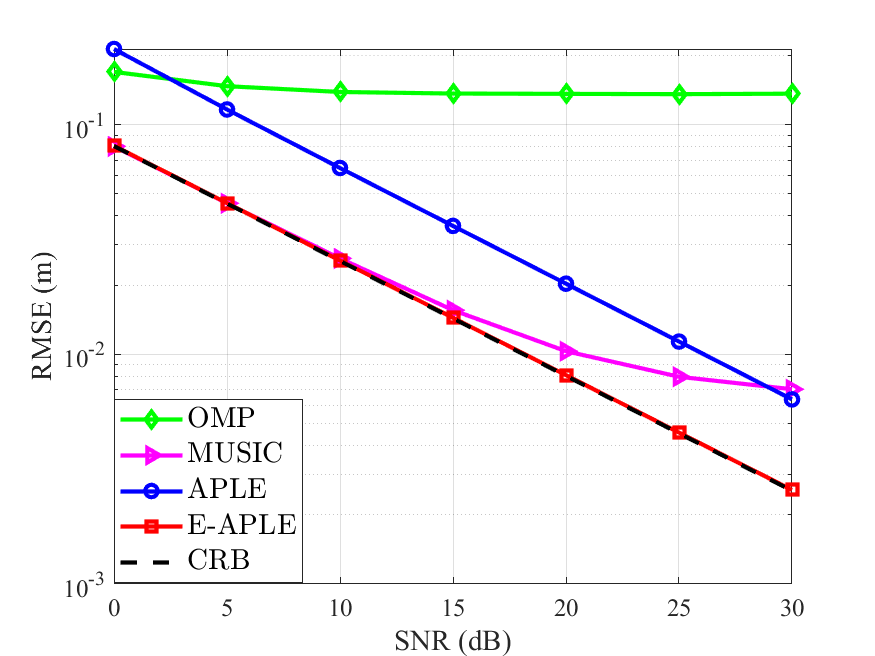}
		\caption{$r=3$ m}
		\label{3m}
	\end{subfigure}
 	\begin{subfigure}{0.9\linewidth}
		\centering
		\includegraphics[width=1\linewidth]{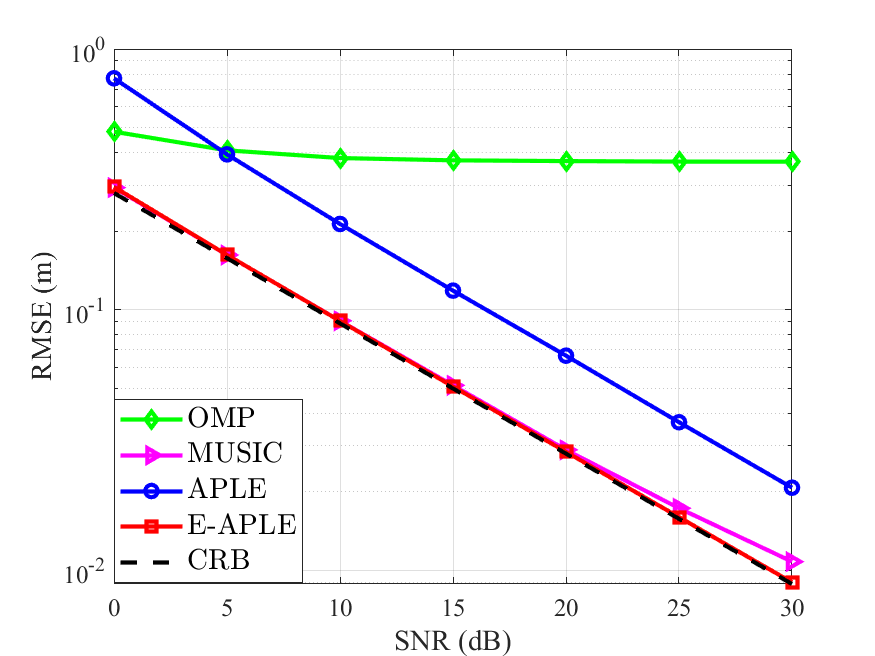}
		\caption{$r=5$ m}
		\label{5m}
	\end{subfigure}
	\caption{The RMSE v.s. SNR for the E-APLE and APLE algorithms and their baselines with $r=1$ m, $r=3$ m and $r= 5$ m.}
	\label{SNR}
\end{figure}
\section{Numerical Results}\label{S7}
In this section, we numerically evaluate the performance of the proposed APLE and E-APLE algorithms. For comparison, we introduce two baselines. For the first baseline, we extend the orthgonal matching pursuit (OMP) algorithm in \cite{OMP9598863} to the three-dimensional case. We first construct a sensing matrix $\mathbf{W}$ in the distance-angle polar domain, where the columns of $\mathbf{W}$ are near-field steering vectors with the sampled distance and angles. The grid resolutions for the distance $r$ and the angles $(\omega,\phi)$ are $0.1$ m and $0.02$ rad, respectively. We search for the column of $\mathbf{W}$ that is most correlated with the received signal to obtain the distance and angle estimates. The extended algorithm is referred to as ``OMP''. 
For the second baseline, we  employ a strategy that separates the estimation of the orientation angles and the distance to the UE. Specifically, we follow the method outlined in \cite{CSNFC9709801} to construct a special covariance matrix that contains only the angle information of the UE location. 
The construction of this covariance matrix is based on a simplified channel model, adopting the Fresnel approximation \cite{Fresappro}.
Then, we employ the MUSIC algorithm to estimate the angles $(\omega,\phi)$ from this covariance matrix \cite{OMUSIC}. With the angle estimates fixed, we again apply the MUSIC algorithm to the received signal for an estimate of $r$. The resulting algorithm is referred to as ``MUSIC''. The CRB derived in \eqref{CRB} serves as the fundamental performance lower bound of all the considered algorithms. Additionally, the MCRB derived in \eqref{FMCRB} is used as the performance lower bound of the APLE algorithm to accommodate the model mismatch introduced by the SFM. 
\par The performance of the proposed APLE and E-APLE algorithms are measured by the root mean-square-error (RMSE) of the estimate of $\mathbf{p}_{\mathrm{U}}$, defined as
\begin{equation}\label{RMSE}
   \operatorname{RMSE}\left(\mathbf{p}_{\mathrm{U}}\right) \triangleq \sqrt{\mathbb{E}\left[{\left\|\mathbf{p}_{\mathrm{U}}-\hat{\mathbf{p}}_{\mathrm{U}}\right\|^{2}}\right]} . 
\end{equation}
In simulations, the expectation in \eqref{RMSE} is numerically approximated by averaging the results of 500 Monte Carlo random experiments. For simplicity, we assume that the BS array and the subarrays are of square shapes with $ N_{\mathrm{x}}= N_{\mathrm{y}}$ and $ N_{m,\mathrm{x}}= N_{m,\mathrm{y}}$, $m\in \mathcal{I}_{M}$. The antenna spacing at the BS array is set to $d_{\mathrm{x}}=d_{\mathrm{y}}=d$. The carrier frequency is set to $f=10$ GHz, corresponding to a wavelength of $\lambda = 0.03$ m. We generate the UE location ${\mathbf{p}}_{\mathrm{U}}$ by specifying its distance $r$ and angle parameters $\omega$ and $\phi$. Unless otherwise stated, the azimuth and polar angles of the UE are randomly drawn from $\omega \in [0,2\pi)$ and $\phi \in [0,\pi/2)$, respectively. 

In Fig. \ref{SNR}, we evaluate the performance of APLE, E-APLE, and baseline methods under a varying SNR, ranging from $0$ dB to $30$ dB. The BS array size is set to $N_{\mathrm{x}}=N_{\mathrm{y}}=50$. The antenna spacing is set to $d=\frac{\lambda }{4}$. Under this configuration, the Fraunhofer distance of the entire BS array is $18.75$ m. The UE-to-BS distance $r$ is set to $1$ m, $3$ m, and $5$ m in the three subfigures of Fig. \ref{SNR}. For APLE, the BS array is partitioned into $M=5\times 5=25$ subarrays. E-APLE is initialized by the output of APLE. From Fig. \ref{SNR}, we observe that the RMSE by OMP decreases with increasing SNR, but eventually reaches an error floor. This is attributed to the inadequate sampling resolution for both distance and angle. Similarly, the RMSE obtained by MUSIC also suffers from an error floor at high SNR, especially when the UE is close to the BS (e.g., $r=1$ m). This issue arises because the Fresnel approximation employed in MUSIC becomes invalid when $r$ is small. In contrast, the estimation accuracy achieved by E-APLE significantly outperforms other methods and closely approaches the CRB. Specifically, the RMSE obtained by E-APLE is less than $1$ cm for $r=3$ m and $\textrm{SNR}=20$ dB. Meanwhile, the RMSE yielded by APLE exhibits a performance gap compared to the CRB due to the model mismatch introduced in the SFM.

\begin{figure}[t]
		\centering
		\includegraphics[width=1\linewidth]{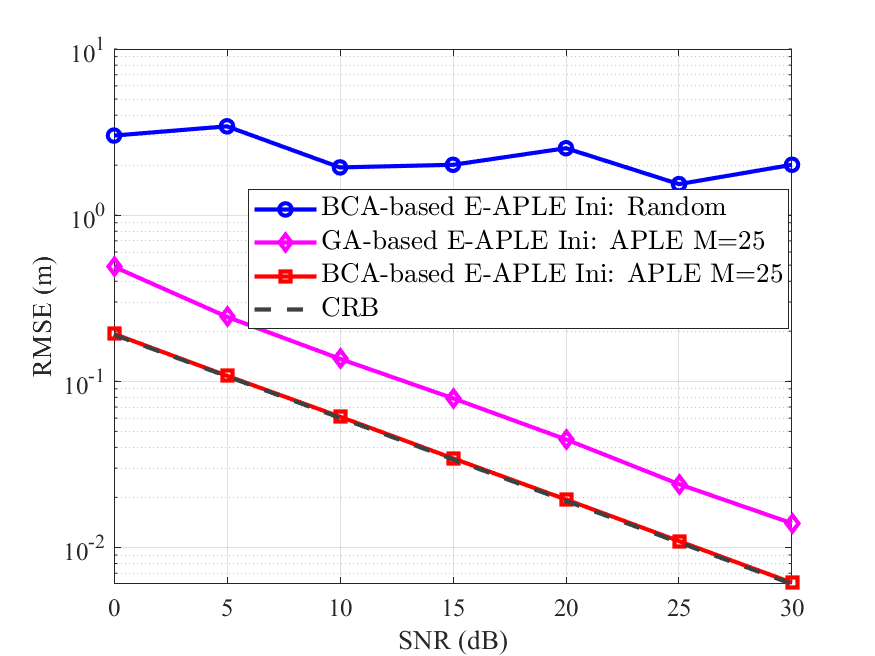}
	\caption{The RMSE curves of the E-APLE algorithm and the CRB against different initialization and optimization strategies. }
	\label{EAPLE}
\end{figure}

Fig. \ref{EAPLE} illustrates the impact of different initialization and optimization strategies on the RMSE performance of the E-APLE algorithm. We set $d=\frac{\lambda }{2}$ and $N_{\mathrm{x}}=N_{\mathrm{y}}=60$. The UE-to-BS distance $r$ is randomly drawn from the
range $[9,11]$ m. We first compare the performance by E-APLE with different initialization strategies. E-APLE is initialized either by the output of APLE with $M=25$ or by a randomly generated location. For the latter, we randomly drawn the UE-to-BS distance and orientation angles from $r\in[9,11]$ m, $\omega\in[0,2\pi)$, and $\phi\in[0,\pi/2)$ to generate a starting point.
From Fig. \ref{EAPLE}, it is observed that the estimation performance of E-APLE initialized by the output of APLE is significantly better than that initialized by a random location. This demonstrates the sensitivity of E-APLE's performance to initialization, with the output of APLE providing a promising starting point. Additionally, we provide the performance of directly using the GA method to solve the ML problem \eqref{FinalNLS}, referred to as ``GA-based E-APLE''. The GA-based E-APLE is also initialized by the output of APLE. It is observed that the estimation error performance of the proposed BCA-based E-APLE clearly outperforms that of GA-based E-APLE.

\begin{figure}[t]
		\centering
		\includegraphics[width=1\linewidth]{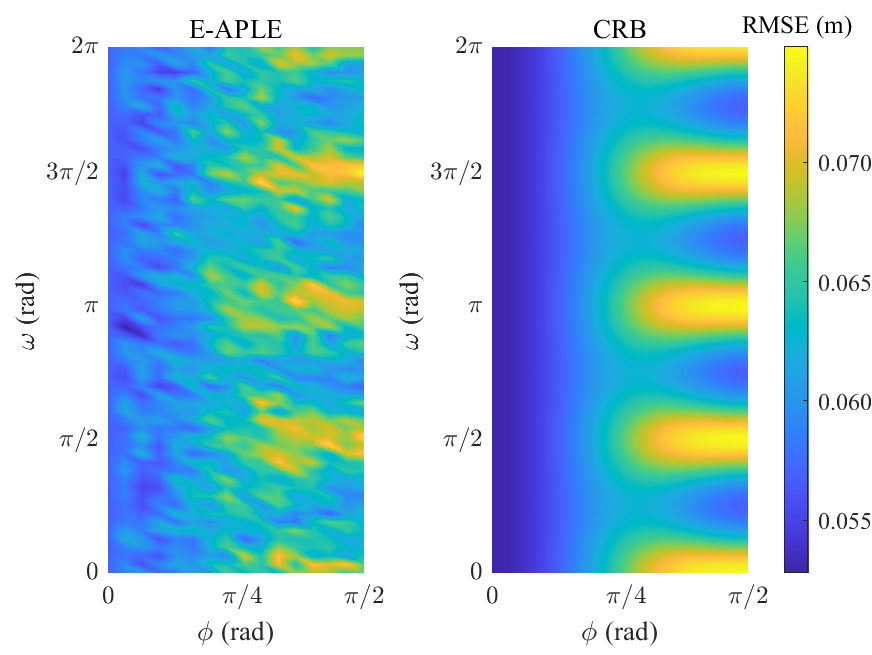}
	\caption{The RMSE of the E-APLE algorithm and the CRB against different azimuth angles $\omega$ and polar angles $\phi$ of the UE. $\omega \in [0,2\pi)$ and $\phi \in [0,\pi/2)$. }
	\label{Angle}
\end{figure}

Fig. \ref{Angle} illustrates the RMSE performance of the E-APLE algorithm and the CRB under varying UE orientations. We set $d=\frac{\lambda }{2}$ and $N_{\mathrm{x}}=N_{\mathrm{y}}=60$. The UE-to-BS distance is fixed to $r=10$ m. The azimuth and the polar angles $\omega$ and $\phi$ are randomly drawn from $\omega \in [0, 2\pi)$ and $\phi \in [0, \pi/2)$. From Fig. \ref{Angle}, it is observed that the performance of the E-ALPLE and the CRB deteriorates when $\omega$ approaches $0$, $\frac{\pi}{2}$, $\pi$, and $\frac{3\pi}{2}$, and when $\phi$ approaches $\frac{\pi}{2}$. This is because, with these values of $\omega$ and $\phi$, the UE is located close to the $xOy$ plane, and the projection of vector $\mathbf{p}_{\mathrm{U}}$ onto the $xOy$ plane is parallel to the $x$-axis or $y$-axis. As a result, the received signal contains less distance and orientation information of the UE location, leading to a decrease in the estimation accuracy of the UE location. 

Table~\ref{MAPLE} illustrates the RMSE performance of the APLE algorithm under varying numbers of BS subarrays. We set $d=\frac{\lambda }{2}$, SNR $=20$ dB, and fixed the UE-to-BS distance to $r=20$ m. The azimuth and polar angles $\omega$ and $\phi$ are randomly drawn from $\omega \in [0, 2\pi)$ and $\phi \in [0, \pi/2)$. The BS array sizes $N_{\mathrm{x}}$ and $N_{\mathrm{y}}$ are set to $60$, $90$, and $120$, and the number of BS subarrays $M$ is set to $4$, $9$, and $16$. We also provide the MCRBs for different array sizes and values of $M$ as the baseline. From Table~\ref{MAPLE}, we observe that the APLE can closely approach the MCRB under various settings. The larger the array size, the better the RMSE performance by the APLE algorithm. For a fixed array size, a smaller $M$ leads to better estimation accuracy. This is because the increase in $M$ weakens the AoA drifting effect between different subarrays.

\begin{table*}[ht]
    \centering
     \caption{The RMSE of the APLE algorithm and the MCRB against $M$ for different $N_\mathrm{x}\times N_\mathrm{y}$.}
     \label{MAPLE}
        \begin{tabular}{ccccccc}
            \toprule
             & \multicolumn{3}{c}{\textbf{APLE} }& \multicolumn{3}{c}{\textbf{MCRB} } \\
            \cmidrule(lr){2-4} \cmidrule(lr){5-7} 
            \textbf{$N_\mathrm{x}\times N_\mathrm{y}$} & $M= 4$ & $M= 9$  & $M= 25$  & $M= 4$ & $M= 9$  & $M= 25$    \\
            \midrule
$60\times60$ & 0.2915 & 0.3960 &  0.6345 & 0.2851 & 0.3940 & 0.6339  \\
$90\times90$ & 0.0873 & 0.1169 & 0.1885 & 0.0843 & 0.1163 & 0.1869\\
$120\times120$ & 0.0385 & 0.0513 & 0.0820 & 0.0365 & 0.0507 & 0.0816\\

            \bottomrule
        \end{tabular}
\end{table*}  

\begin{table*}[ht]
    \centering
     \caption{The RMSEs of the APLE and E-APLE algorithms and the CRB against $r$ for different $N_\mathrm{x}\times N_\mathrm{y}$.}
     \label{REAPLE}
        \begin{tabular}{cccccccccc}
            \toprule
             & \multicolumn{3}{c}{\textbf{APLE} }& \multicolumn{3}{c}{\textbf{E-APLE} }& \multicolumn{3}{c}{\textbf{CRB} } \\
            \cmidrule(lr){2-4} \cmidrule(lr){5-7} \cmidrule(lr){8-10} 
            \textbf{$N_\mathrm{x}\times N_\mathrm{y}$} & $r= 10$ m & $r= 20$ m & $r=30$ m & $r= 10$ m & $r= 20$ m & $r=30$ m & $r= 10$ m & $r= 20$ m & $r=30$ m   \\
            \midrule
$60\times60$ & 0.0442 & 0.1647 & 0.3629 & 0.0174 & 0.0663 & 0.1467 & 0.0171 & 0.0661 &  0.1403 \\
$90\times90$ & 0.0141 & 0.0492 & 0.1137 & 0.0048 & 0.0172 & 0.0380 & 0.0045 & 0.0172 & 0.0370 \\
$120\times120$ & 0.0062 & 0.0218 & 0.0488 &  0.0022 & 0.0085 & 0.0183 &  0.0022 & 0.0084 &  0.0183\\

            \bottomrule
        \end{tabular}
\end{table*}

\begin{table*}[t]
    \centering
     \caption{The runtimes and RMSEs of the E-APLE and APLE algorithms and their baselines with various $N_\mathrm{x}\times N_\mathrm{y}$.}
     \label{Runtime}
        \begin{tabular}{ccccccccccc}
            \toprule
             & \multicolumn{4}{c}{\textbf{Runtime (s)}} & \multicolumn{6}{c}{\textbf{ RMSE (m) }} \\
            \cmidrule(lr){2-5} \cmidrule(lr){6-11} 
            \textbf{$N_\mathrm{x}\times N_\mathrm{y}$} & APLE & E-APLE & MUSIC &  OMP & APLE & E-APLE & MUSIC &  OMP & MCRB & CRB \\
            \midrule
            $50\times50$ & 0.0069 & 1.5667 & 3.8485 &  12.6252 & 0.1220 & 0.1102 & 0.1107 & 0.3618 & 0.1208 & 0.1091 \\

            $75\times75$ & 0.0135 & 6.7765 & 29.8928 &  29.1471 & 0.0504 & 0.0331 & 0.0335 & 0.3496 &  0.0495 &  0.0323 \\

            $100\times100$ & 0.0236 & 19.4468 & -   & - & 0.0275 & 0.0137 &  - & - & 0.0273 & 0.0137\\

            \bottomrule
        \end{tabular}
\end{table*}  

Table~\ref{REAPLE} shows the RMSE performance of the APLE and E-APLE algorithms under varying UE-to-BS distances. We set $N_{\mathrm{x}}$ and $N_{\mathrm{y}}$ to $60$, $90$, and $120$, with $d=\frac{\lambda }{2}$ and SNR $=20$ dB. The UE-to-BS distance $r$ is varied as $10$ m, $20$ m, and $30$ m. The CRBs under different array sizes and distances are also provided as the baseline for the APLE and the E-APLE algorithms. From Table~\ref{REAPLE}, it is clear that for a fixed $r$, larger array sizes result in better RMSE performance for both the APLE and E-APLE algorithms. The RMSE performance of APLE is slightly inferior to that of E-APLE. Under various distance settings, the RMSE of E-APLE can closely approach the CRBs. Under the setting of $N_{\mathrm{x}}=N_{\mathrm{y}}=120$ and $r=20$ m, the RMSE by E-APLE is less than $1$ cm. 

 Table~\ref{Runtime} presents the RMSE performance and runtimes of compared methods under varying BS array sizes. The experiments are conducted on MATLAB R2020b on a Windows x64 computer with a 3 GHz CPU and 64 GB RAM. We set $d=\frac{\lambda }{4}$, SNR $=20$ dB, and $r = 10$ m. $N_{\mathrm{x}}=N_{\mathrm{y}}$ is varied as $50$, $75$, and $100$, and the BS array is partitioned into $4$, $9$, and $16$ subarrays, respectively, for the APLE algorithm. In Table~\ref{Runtime}, for $ N_{\mathrm{x}} \times N_{\mathrm{y}}=100 \times 100$, both the OMP and MUSIC algorithms run out of memory. We observe that APLE and E-APLE exhibit significant runtime advantages over the baseline methods, especially with larger BS array sizes. The runtimes of APLE and E-APLE scale nearly linearly with the number of BS antennas. This validates the scalability of APLE and E-APLE in ELAA scenarios. Furthermore, in terms of RMSE performance, E-APLE outperforms the baseline methods.

\section{Conclusions}\label{S8}
In this paper, we investigated the 3D near-field UE location estimation problem in the SIMO system. By partitioning the BS array into multiple subarrays, the UE can be regarded as located in the far-field region of each subarray. We established a probabilistic model for UE location estimation by leveraging the geometric correlations between the AoA at each subarray and the UE location. We proposed a low-complexity scalable UE location estimation algorithm, namely, APLE, by using message-passing techniques. We further developed the E-APLE algorithm to improve the localization accuracy by following the ML principle. Numerical results demonstrated that the APLE and the E-APLE algorithms achieve remarkable localization accuracy and exhibit excellent scalability as the array size goes large.

\appendices 
\section{Derivations of \eqref{20_1} and \eqref{20_2}}
\label{appA}
For simplicity, we denote $\mu _{\theta_{n,v}\rightarrow \varphi_{n,v}}$ and $ \kappa _{\theta_{n,v}\rightarrow \varphi_{n,v}}$ by $\mu_{n,v}$ and $\kappa_{n,v}$, respectively. The expression of $\Delta _{\mathbf{p}_{\mathrm{U}}\rightarrow\varphi_{m,u} }$ is given by
 \begin{multline}
 \Delta _{\mathbf{p}_{\mathrm{U}}\rightarrow\varphi_{m,u} } \propto \\
 \exp{\left\{\sum_{(n,v)\in\mathcal{A}\setminus(m,u)}{ \kappa_{n,v}\cos\left(\pi \mathbf{e}^{\mathrm{T}}_{n}\mathbf{e}_{v} -{\mu} _{n,v} \right)}\right\}} \label{appdelta}   
 \end{multline}
with $\theta_{n,v}=\mathbf{e}^{\mathrm{T}}_{n}\mathbf{e}_{v}$ and $\mathbf{e}_{n} = \frac{\mathbf{p}_\mathrm{U}-\mathbf{p}_{\mathrm{BS},n}}{\|\mathbf{p}_\mathrm{U}-\mathbf{p}_{\mathrm{BS},n}\|}$. We resort to the GA method to find the local maxima of \eqref{appdelta}, which serves as the mean vector $\mathbf{m}_{\mathbf{p}_{\mathrm{U}}\rightarrow\varphi_{m,u}}$ of the approximated Gaussian message in \eqref{15}. The covariance matrix $\mathbf{C}_{\mathbf{p}_{\mathrm{U}}\rightarrow\varphi_{m,u}}$ of the approximated Gaussian message is given by the Hessian matrix at $\mathbf{m}_{\mathbf{p}_{\mathrm{U}}\rightarrow\varphi_{m,u}}$. Denote by $\varpi_{m,u}(\mathbf{p}_{\mathrm{U}})$ the exponential term of \eqref{appdelta}. The gradient of $\varpi_{m,u}(\mathbf{p}_{\mathrm{U}})$ with respect to $\mathbf{p}_{\mathrm{U}}$ is
	\begin{multline}
		 \nabla \varpi_{m,u}(\mathbf{p}_{\mathrm{U}})=\\
  \sum_{(n,v)\in\mathcal{A}\setminus(m,u)}{\left(-\pi\kappa_{n,v}\sin\left(\pi\mathbf{e}^{\mathrm{T}}_{n}\mathbf{e}_{v}-{\mu} _{n,v}\right)\mathbf{u}_{n,v}\right)},
	\end{multline}
with $\mathbf{u}_{n,v}=\frac{\mathbf{e}_v-{\mathbf{e}^\mathrm{T}_{n}}\mathbf{e}_v\mathbf{e}_{n}}{\left\|\mathbf{p}_\mathrm{U}-\mathbf{p}_{\mathrm{BS},n} \right\|}$. 
Denote by $\mathbf{p}^{(\mathrm{old})}_{\mathrm{U}}$ the $\mathbf{p}_{\mathrm{U}}$ obtained in the previous iteration. The update of $\mathbf{p}_{\mathrm{U}}$ in the current iteration is given by
\begin{equation}
\mathbf{p}^{(\mathrm{new})}_{\mathrm{U}}=\mathbf{p}^{(\mathrm{old})}_{\mathrm{U}} + \epsilon_3 \left.  \nabla \varpi_{m,u}(\mathbf{p}_{\mathrm{U}}) \right|_{\mathbf{p}_{\mathrm{U}}=\mathbf{p}^{(\mathrm{old})}_{\mathrm{U}}},
\end{equation}
 where $\epsilon_3>0$ is an appropriate step size that can be selected from the backtracking line search to satisfy
\begin{equation}
  \varpi_{m,u}\left(\mathbf{p}^{(\mathrm{new})}_{\mathrm{U}}\right) \geq   \varpi_{m,u}\left(\mathbf{p}^{(\mathrm{old})}_{\mathrm{U}}\right).
\end{equation}
An update of $\mathbf{p}_{\mathrm{U}}$ is obtained after $T_p$ iterations. $\mathbf{m}_{\mathbf{p}_{\mathrm{U}}\rightarrow\varphi_{m,u}}$ is set by the obtained local optima $\hat{\mathbf{p}}_{\mathrm{U},(m,u)}$.
For the calculation of $\mathbf{C}_{\mathbf{p}_{\mathrm{U}}\rightarrow\varphi_{m,u}}$, the Hessian matrix of $\varpi_{m,u}(\mathbf{p}_{\mathrm{U}})$ is expressed as
\begin{align}
\mathbf{H}(\mathbf{p}_{\mathrm{U}}) =& \sum_{(n,v)\in\mathcal{A}\setminus(m,u)}{\left(-\pi^2\kappa_{n,v}\cos\left(\pi\mathbf{e}_{n}^{\mathrm{T}}\mathbf{e}_{v}-\mu_{n,v}\right)\right.}\notag\\
	&{\left.\times\mathbf{u}_{n,v}\mathbf{u}_{n,v}^{\mathrm{T}}-\pi\kappa_{n,v}\sin\left(\pi\mathbf{e}_{n}^{\mathrm{T}}\mathbf{e}_{v}-\mu_{n,v}\right)\right.}\notag\\
 &{\left.\!\times\left(\!\frac{3\mathbf{e}_{n}^\mathrm{T}\mathbf{e}_v\mathbf{e}_{n}{\mathbf{e}_{n}}^\mathrm{T}}{\left\| \mathbf{p}_\mathrm{U}-\mathbf{p}_{\mathrm{BS},n} \right\|^2} \!- \!\frac{ \mathbf{e}_{n}^\mathrm{T}\mathbf{e}_v\mathbf{I}_3 + \mathbf{e}_{n}\mathbf{e}_v^\mathrm{T} + \mathbf{e}_v\mathbf{e}_{n}^\mathrm{T} }{\left\| \mathbf{p}_\mathrm{U}-\mathbf{p}_{\mathrm{BS},n} \right\|^2}\right) \!\!\right)}.
\end{align}     
Then, we approximate $\mathbf{C}_{\mathbf{p}_{\mathrm{U}}\rightarrow\varphi_{m,u}}$ as
\begin{equation}
	\mathbf{C}_{\mathbf{p}_{\mathrm{U}}\rightarrow\varphi_{m,u}} = \left(-\left.\mathbf{H}(\mathbf{p}_{\mathrm{U}})\right|_{\mathbf{p}_{\mathrm{U}}=\hat{\mathbf{p}}_{\mathrm{U},(m,u)}}
\right)^{-1}.
\end{equation}

\bibliographystyle{IEEEtran}
\bibliography{ScalableNearField}
\end{sloppypar}
\end{document}